\documentclass[12pt]{amsart}
\usepackage{graphicx}
\usepackage{xcolor}
\usepackage{pifont}
\usepackage{placeins}
\usepackage[bookmarksnumbered,colorlinks,bookmarks,citecolor=blue,%
	linkcolor=blue,urlcolor=blue,breaklinks,linktocpage]{hyperref}
\usepackage[all]{hypcap}

\newcommand{\empar}[1]{\medskip\noindent{\bf #1.}\ }

\def\boldit#1{\textit{\textbf{#1}}}

\def\plotsym{$\sss\bullet$}
\def\Narc{\beginpicture
    \multiput{} at -1 -1  1 1 /
    \plotsymbolspacing=1pt
    \setplotsymbol({\plotsym})
    \setquadratic
    \plot -.5 .3  0 .5  .5 .3 /
    \endpicture
    }
    
\def\Ntri{\beginpicture
    \multiput{} at -1 -1  1 1 /
    \plot -.6 0  .6 0 /
    \plot -.7 .3  -.3 .7 /
    \plot .3 .7  .7 .3 /
    \endpicture
    }
    
 \def\arr#1#2{\arrow <2mm> [0.25,0.75] from #1 to #2}

\def\Sarc{\beginpicture
    \multiput{} at -1 -1  1 1 /
    \plotsymbolspacing=1pt
    \setplotsymbol({\plotsym})
    \setquadratic
    \plot -.5 -.3  0 -.5  .5 -.3 /
    \endpicture
    }
    
\def\SWarc{\beginpicture
    \multiput{} at -1 -1  1 1 /
    \plotsymbolspacing=1pt
    \setplotsymbol({\plotsym})
    \setquadratic
    \plot -.9 .5  -.3 -.3  .5 -.9 /
    \endpicture
    }
    
\def\NWarc{\beginpicture
    \multiput{} at -1 -1  1 1 /
    \plotsymbolspacing=1pt
    \setplotsymbol({\plotsym})
    \setquadratic
    \plot -.9 -.5  -.3 .3  .5 .9 /
    \endpicture
    }
    
\def\NEarc{\beginpicture
    \multiput{} at -1 -1  1 1 /
    \plotsymbolspacing=1pt
    \setplotsymbol({\plotsym})
    \setquadratic
    \plot -.5 .9  .3 .3  .9 -.5 /
    \endpicture
    }
    
\def\SEarc{\beginpicture
    \multiput{} at -1 -1  1 1 /
    \plotsymbolspacing=1pt
    \setplotsymbol({\plotsym})
    \setquadratic
    \plot -.5 -.9  .3 -.3  .9 .5 /
    \endpicture
    }

\input prepictex   \input pictex    
\input postpictex
  \newcommand\F{{\sf F}}
  \newcommand\A{{\sf A}}
  \newcommand\C{{\sf C}}
  \newcommand\E{{\sf E}}
  \newcommand\G{{\sf G}}
  \newcommand\B{{\sf B}}
  \newcommand\D{{\sf D}}
  \newcommand\p{$'$}
  \renewcommand\c{$_\prime$}
  \newcommand\s{$\sharp$}
  \newcommand\f{$\flat$}
  \newcommand\V{$\,\check{}$}
  \renewcommand\H{$\,\hat{}$}

\begin{document}
  
  \begin{center}
    {\large\bf 2:3:4 Harmony within the Tritave}

    \bigskip
    {\normalsize Markus Schmidmeier\footnote{This work was partially supported by the
        Simons Foundation (Grant number 245848).}}

 \bigskip

  \begin{minipage}{12cm}
    \footnotesize
  {\bf Abstract:}   In the Pythagorean tuning system, the fifth is used to generate
  a scale of 12 notes per octave.
  In this paper, we use the octave to generate
  a scale of 19 notes per tritave; one can play this scale
  on a traditional piano.  In this system,
  the octave becomes a proper
  interval and the 2:3:4 chord a proper chord.
  We study harmonic properties obtained from the 2:3:4 chord,
  in particular composition elements using 
  dominants, inversions, major and minor chords, and diminished chords.
  The {\it Tonnetz} (array notation)
  turns out to be an effective tool to visualize
  the harmonic development in a composition based on these elements.
  2:3:4 harmony may sound
  pure, yet sparse, as we illustrate in a short piece.


\end{minipage}
  \end{center}

\section*{Introduction}

The 4:5:6 chord lies at the very center of major mode harmony
in the music of the western world.  It is ubiquitous in chord sequences
involving dominants and subdominants, and its mirror image
defines the associated minor chords.
Allowing for inversion, the chord appears in many shapes including the
3:4:5 chord with the stunning harmonicity given by its low frequency ratios.

By comparison the 2:3:4 chord, despite being 
admired since Pythagoras' time for its perhaps unsurpassable purity, 
appears underused in tonal music.
In fact, traditional harmony may not consider the 2:3:4 chord as being proper
since chords are defined based on the pitch classes of their notes, up to octave equivalence.

In this paper we take the position
that it is the chord, or the chord sequence, which defines what is harmony.
If octave equivalence does not fit with the chord, then we have to replace it.
This has been done before:  the Bohlen-Pierce scale,
see \cite{advocat}, is based on the 3:5:7 chord with equivalence modulo
the {\it tritave} or {\it duodecime}, the interval given by the
frequency ratio 3:1.  There, octave-equivalence is 
not even considered a viable concept.

The system which we propose to study is based on the 2:3:4 chord, modulo tritave-equivalence.
Thus we offer an interesting alternative to the traditional octave-equivalence-based triadic system.

\bigskip
Our paper has three parts and some appendices.
We discuss the third part first.
Here, we take elements of composition from the 4:5:6 system
and adapt them to 2:3:4 harmony.

Recall that the dominant of a chord in the 4:5:6 system is formed by replacing
each note by its successor in the circle of fifths, and by adjusting with first and
second inversions, which move the base note of a chord up by one octave,
or the top note down, respectively.
For 2:3:4 harmony, there is a corresponding concept, the {\it circle
of octaves,} which we introduce in Section~\ref{subsection-circle-of-octaves}.
First and second inversions in a tritave-based system are
obtained by moving the base note up by one tritave, or the top note down.
In each system, for the subdominant, one proceeds in the opposite direction
in the circle.  Together, 
tonic, dominant and subdominant form
what we call in Section~\ref{subsection-dominant} the {\it basic sequence,}

\begin{center}
tonic --- subdominant --- dominant --- tonic,
\end{center}

which we picture first in the 4:5:6 system,
starting with the \C-\E-\G\ chord:

\begin{center}
\includegraphics{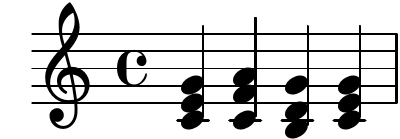}
\end{center}

A sample 2:3:4 chord is \A-\E-\A\p\ as the \E\ is a fifth (3:2) above the \A, and the \A\p\ an
octave (4:2) above the \A.  Here is the basic sequence for the \A-\E-\A\p\ chord:

\begin{center}
\includegraphics{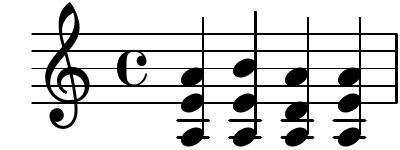}
\end{center}

Namely, starting from the 2:3:4 chord \A-\E-\A\p, the second (subdominant) chord
is obtained as follows.  Replace each note in \A-\E-\A\p\ by the note one
octave lower, this yields \A\c-\E\c-\A.  Since  \A\c\ and \E\ are one tritave
apart, and so are \E\c\ and \B\p, we obtain the chord \A-\E-\B\p,
which is {\it tritave-equivalent} to \A\c-\E\c-\A.
For the dominant, we move each note in \A-\E-\A\p\ up by an octave to get
\A\p-\E\p-\A\p\p, which is tritave-equivalent to \A-\D-\A\p.

We will see in Section~\ref{section-inversions} that the role of
{\it major and minor chords} is formally and audibly quite different.
In Section~\ref{subsection-diminished} we study the use of
{\it diminished chords} for the development of the harmony.

\sloppypar
Throughout the third part of the paper, we refer to a sample piece
{\it Ave Maria in dix-neuf par duodecime} which has been composed by the author.
While the harmony explores the above discussed
elements of composition in 2:3:4 harmony,
the arrangement follows loosely J.\ S.\ Bach's C-major {\it Prelude} from WTK1.
We reprint the score in Appendix~\ref{appendix-ave};
a recording is available
on YouTube at {\tt \footnotesize https://youtu.be/Bg1n4jM1n5w}.

We discuss perception of music in the 2:3:4 system in the final
Section~\ref{subsection-purity} on {\it purity and sparsity.}

\bigskip\noindent
In the first part of this paper,
we construct and analyze a Pythagorean scale of 19 notes per tritave
in which the 2:3:4 chord can be played and modulated.
Harmonically, the notes are arranged in the {\it circle of octaves,}
which we introduce and discuss in Section~\ref{subsection-circle-of-octaves}.

A nice feature of this scale is that in practice,
it can be played on a traditional piano, either in just Pythagorean
or in 12 tone equal intonation, as we will see in
Section~\ref{subsection-piano}.

Nevertheless, as the scale is tritave-based,
a keyboard which emphasizes periodicity with respect to the
tritave would be preferable.

\medskip
\begin{center}
\setlength\unitlength{.14in}
\begin{picture}(33,8)(0,-2.5)
\put(-.5,0){\line(1,0){34}}
\multiput(0,0)(1,0){34}{\line(0,1){5}}
\linethickness{.07in}
\multiput(0,0)(11,0)3{\multiput(5,1.6)(1,0)2{\line(0,1){3.4}}}
\multiput(0,0)(11,0)3{\multiput(1,1.6)(1,0)3{\line(0,1){3.4}}}
\multiput(0,0)(11,0)3{\multiput(8,1.6)(1,0)3{\line(0,1){3.4}}}
\put(5,-1.5){{\G\c\c}}
\put(16,-1.5){{\D}}
\put(27,-1.5){{\A\p\p}}
\end{picture}
\end{center}

In the second part of this paper,
we review a method for visualizing harmony, the {\it Tonnetz,}
and adapt it for 2:3:4 harmony.  The {\it Tonnetz} captures
the above mentioned elements of composition; sequences of triangles or arcs
within the picture visualize how harmony develops throughout a
piece.

In the 4:5:6 {\it Tonnetz,} there are major thirds (5:4) along the up-diagonal,
minor thirds (6:5) along the down-diagonal, and hence fifths (6:4) in horizontal direction.
By comparison, the 2:3:4 {\it Tonnetz} has fifths (3:2) along the up-diagonal, fourths (4:3) along the
down-diagonal and octaves (4:2) in horizontal direction.

In Section~\ref{section-neo-Riemannian} we adapt the PLR-moves from neo-Riemannian theory
to the 2:3:4 {\it Tonnetz.}
While the moves capture elements of 2:3:4 harmony quite well,
it turns out that, when compared to 4:5:6 harmony, the number of notes that can be reached with
a given number of PLR-moves is smaller.

\section{A piano scale based on the tritave}

The famous octave-based Pythagorean scale turns out to have a
lesser known relative, the tritave-based Pythagorean scale Pyth-3.
We briefly review the Pythagorean scale Pyth-2
in order to introduce Pyth-3.

In each Pythagorean scale, a note is given by a relative frequency of the form
$2^u3^v$ where the harmonic degree $v$ describes the position of the note
in Pyth-2 in the circle of fifths.  The corresponding arrangement of notes
by harmonic degree in Pyth-3 is in the circle of octaves; here the
position of a note is given by $u$.  Finally, we discuss in this section the
playability of Pyth-3 on a traditional piano.

More comprehensive reviews of the
Pythagorean scale Pyth-2 can be found
in the literature, see for example \cite{liern}.
For the construction of well-formed scales and a discussion of
desired properties like closure and symmetry we refer to
\cite{cc}, see Appendix~\ref{appendix-a}.

\subsection{Two harmonic degrees for Pythagorean scales}
In order to describe notes in Pythagorean scales, two
ingredients are needed:
\begin{itemize}
\item A base note, which we call \D.
\item Two intervals, the octave (2:1) and the tritave (3:1).
\end{itemize}
An arbitrary note in the scale is obtained from the base note by going
up a certain number $u$ of octaves and a certain number $v$ of tritaves.  Either
or both numbers can be negative.  We denote by
\begin{itemize}
\item Pyth the collection of all notes obtained in this way.
\end{itemize}
One can also use the octave and the fifth (3:2) for the construction,
but the scale and the set Pyth will be the same.

To each note one assigns a frequency ratio $\varphi$,
it is the quotient of the frequency of the given note by the frequency 
assigned to the \D.  This number can be writen as
$$\varphi = 2^u \cdot 3^v$$
where $u=\mu_2(\varphi)$ is the 2-adic valuation 
and $v=\mu_3(\varphi)$ the 3-adic valuation of $\varphi$.
Thus, the frequency ratios of the notes in Pyth form the subgroup
of the multiplicative group $\mathbb Q^+$ generated by 2 and 3.

The numbers $\mu_2(\varphi)$ and $\mu_3(\varphi)$ are called the
{\it harmonic degrees} of the note given by frequency ratio $\varphi$.
We single out two subsets of Pyth:

\begin{itemize}
\item Pyth-2 consists of all notes in Pyth of harmonic degree $\mu_3$
      between $-5$ and $+6$.
\item Pyth-3 consists of all notes in Pyth of harmonic degree $\mu_2$
      between $-9$ and $+9$.
\end{itemize}

The bounds of the harmonic degrees will become transparent in the next section.

\subsection{The comma and enharmonic notes}
\label{subsection-enharmonic}
So far, we have infinitely many notes, in Pyth there is one for each pair $(u,v)$ of
integers.  To reduce this number, certain notes are considered
to be ``enharmonic''.  We show that each note is enharmonic to
a unique one in Pyth-2 and to a unique one in Pyth-3.

It turns out that the numbers $2^{19}$ and $3^{12}$ are very close,
their quotient
$$\kappa=\frac{3^{12}}{2^{19}}=1.01364\ldots$$
is the {\it Pythagorean comma.}  In cents it is about $1200\cdot\log_2(\kappa)=23.460\ldots$.

We say that two notes are {\it enharmonic} if their frequency ratio is a power of
the Pythagorean comma $\kappa$.
As a consequence we obtain:

\begin{enumerate}
\item Each note in Pyth is enharmonic to a unique note in Pyth-2.
\item Each note in Pyth is enharmonic to a unique note in Pyth-3.
\end{enumerate}

We explain the second statement for a note of frequency
$\varphi=2^u\cdot 3^v$.  Consider the interval $I=\{-9,-8,\ldots,8,9\}$ of 19 consecutive integers.
By the Division Theorem, there is a unique quotient $m$ and a unique remainder $r\in I$
such that $ u = 19\cdot m+r$.
The note of frequency ratio $\psi=\varphi\cdot \kappa^m$ is enharmonic
with $\varphi$ and has harmonic
degree $\mu_2(\psi)=\mu_2(\varphi\cdot\kappa^m)=u-19\cdot m=r$
which is in the range from $-9$ to $+9$.
Hence $\psi$ is in Pyth-3, and it follows from the uniqueness of $m$ that
$\psi$ is the unique note in Pyth-3 which is enharmonic with $\varphi$.
This shows (2).

So the exponent 19 in the Pythagorean comma $\kappa=3^{12}/ 2^{19}$
determines the number of consecutive integers which can occur
as harmonic degrees in Pyth-3.  Similarly, the exponent 12 determines the
number of harmonic degrees in Pyth-2.

In Appendix~\ref{appendix-number-of-notes}, we describe  
how the numbers 12 and 19 are found, together with other possible pairs.

\subsection{The circle of fifths versus the circle of octaves}
\label{subsection-circle-of-octaves}
For each scale degree, we pick one note and assign it a letter name.
Then any other note of the same scale degree is equivalent,
with respect to the octave or the tritave, to the given note.

The set Pyth-2 is still infinite; the notes with harmonic degree $\mu_3=h$, where $-5\leq h\leq 6$,
have the following frequency ratios:
$$\ldots\quad 2^{-2}\cdot 3^h, \quad 2^{-1}\cdot 3^h, \quad 3^h,
	      \quad 2\cdot 3^h, \quad 2^2\cdot 3^h, \quad \ldots$$
For any two of them, the quotient is a power of two, so
the notes differ by a certain number of octaves.  We say they are
{\it octave-equivalent}.

Fix an interval $I_2=(a,b\,]$ of positive real numbers such that $b/a=2$.
We call this interval $I_2$ a {\it fundamental domain} for Pyth-2.
For the 12 harmonic degrees in Pyth-2, we obtain 12 unique notes in $I_2$;
we list the notes, their frequencies and degrees in
Table~\ref{table-comparison} in the Appendix.
In the circle of fifths, see Figure~\ref{figure-circle-of-fifths}, the notes
are arranged by harmonic degree.

\begin{figure}[ht]
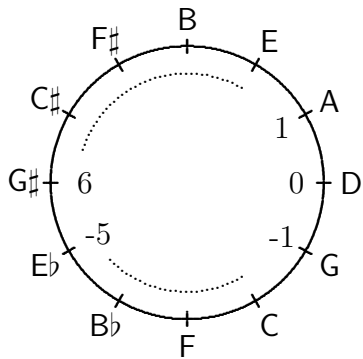

\centerline{$
\beginpicture
\setcoordinatesystem units <1ex,1ex>
\multiput{} at -13 -13 13 13 /
\circulararc 360 degrees from -10 0 center at 0 0
\plot 9.5 0  10.5 0 /            \put {\D} at 12 0
                                 \put 0 at 8 0 
\plot 8.23 4.75  9.09 5.25 /     \put {\A} at 10.4 6
                                 \put 1 at 6.93 4                     
\plot 4.75 8.23  5.25 9.09 /     \put {\E} at 6 10.4
\plot 0 9.5  0 10.5 /            \put {\B} at 0 12
\plot -4.75 8.23  -5.25 9.09 /   \put {\F\s} at -6 10.4
\plot -8.23 4.75  -9.09 5.25 /   \put {\C\s} at -10.4 6
\plot -9.5 0  -10.5 0   /        \put {\G\s} at -12 0 
                                 \put {6} at -7.5 0
\plot -8.25 -4.75  -9.09 -5.25 / \put {\E\f} at -10.4 -6
      	    	   	         \put {-5} at -6.5 -3.75
\plot -4.75 -8.23  -5.25 -9.09 / \put {\B\f} at -6 -10.4
\plot 0 -9.5  0 -10.5  /         \put {\F} at 0 -12
\plot 4.75 -8.23  5.25 -9.09 /   \put {\C} at 6 -10.4
\plot 8.23 -4.75  9.09 -5.25 /   \put {\G} at 10.4 -6
                                 \put {-1} at 6.93 -4
\setdots<2pt>
\circulararc 105 degrees from 4 6.93 center at 0 0
\circulararc -75 degrees from 4 -6.93 center at 0 0 
\endpicture$}
  \caption{The Circle of Fifths in Pyth-2}
  \label{figure-circle-of-fifths}
\end{figure}

We repeat this process for Pyth-3.

Consider the notes with harmonic degree $\mu_2=h$, where $-9\leq h\leq 9$, their frequencies are:
$$\ldots\quad 2^h\cdot 3^{-2}, \quad 2^h\cdot 3^{-1}, \quad 2^h,
	      \quad 2^h\cdot 3, \quad 2^h\cdot 3^2, \quad \ldots$$
For any two of them, the quotient is a power of three, so
the notes differ by a certain number of tritaves.  We say they are
{\it tritave-equivalent.}

Here we take as {\it fundamental domain for Pyth-3}
an interval $I_3=(a,b\,]$ of positive real numbers such that $b/a=3$.
Among the notes in the list,
there is exactly one note with frequency in $I_3$.
For the 19 harmonic degrees in Pyth-3, we obtain 19 unique notes in $I_3$.
If we take in particular the interval
$$I_3=(1/\sqrt 3,\sqrt3\,],$$
we obtain the following notes:

\begin{itemize}
\item the 12 notes \A, $\ldots$, \G\s\ from Pyth-2 in the interval $I_2=(\kappa/\sqrt 2,\kappa\sqrt2\,]$
\item the \A\f\
\item the notes \F\c, \F\s\c, \G\c, which are one
  octave under \F, \F\s, \G, respectively, and
\item the notes \A\p, \B\f\p, \B\p\ one octave above \A, \B\f, \B,
  respectively.
\end{itemize}

In Table~\ref{table-19edt} in the Appendix we list with each note
its frequency, its scale degree, its harmonic degree, and how much it
deviates from equal intonation.

Again, up to rotation and tritave-equivalence, 
the choice of the interval does
not matter, so it is best to arrange the notes in a circle,
see Figure~\ref{figure-circle-of-octaves}.

\begin{figure}[ht]
\centerline{$
\beginpicture
\setcoordinatesystem units <1.2ex,1.2ex>
\multiput{} at -13 -13 13 13 /
\circulararc 360 degrees from -10 0 center at 0 0
\plot 9.5 0  10.5 0 /            \put {\D} at 12 0
                                 \put {0} at 8 0
\plot 8.99 3.09  9.93 3.41 /     \put {\G\c} at 11.35 3.90
                                 \put 1 at 7.57 2.6
\plot 7.50 5.84  8.29 6.65 /     \put {\G} at 9.47 7.37
\plot 5.20 7.95  5.74 8.79 /     \put {\C} at 6.56 10.05
\plot 2.33 9.21  2.58 10.18 /    \put {\F\c} at 2.95 11.63
\plot -.78 9.47  -.87 10.46 /    \put {\F} at -.99 11.96
\plot -3.82 8.7  -4.22 9.62 /    \put {\B\f} at -4.82 10.99
\plot -6.43 6.99  -7.11 7.73 /   \put {\B\f\p} at -8.13 8.83
\plot -8.36 4.52  -9.23 5 /      \put {\E\f} at -10.55 5.71
\plot -9.37 1.56  -10.36 1.73  / \put {\A\f} at -11.84 1.98
                                 \put 9 at -7.8  1.3
\plot -9.37 -1.56  -10.36 -1.73 / \put {\G\s} at -11.84 -1.98
                                 \put {-9} at -7.8 -1.3                         
\plot -8.36 -4.52  -9.23 -5 /    \put {\C\s} at -10.55 -5.71
\plot -6.43 -6.99 -7.11 -7.73 /  \put {\F\s\c} at -8.13 -8.83
\plot -3.82 -8.7  -4.22 -9.62 /  \put {\F\s} at -4.82 -10.99
\plot -.78 -9.47  -.87 -10.46 /  \put {\B} at -.99 -11.96
\plot 2.33 -9.21 2.58 -10.18 /   \put {\B\p} at 2.95 -11.63
\plot 5.20 -7.95  5.74 -8.79 /   \put {\E} at 6.56 -10.05
\plot 7.5 -5.84  8.29 -6.65 /    \put {\A} at 9.47 -7.37
\plot 8.99 -3.09  9.93 -3.41 /   \put {\A\p} at 11.35 -3.9
                                 \put {-1} at 7.57 -2.6
\setdots<2pt>
\circulararc -115 degrees from 6.31 -4.91 center at 0 0
\circulararc 115 degrees from 6.31 4.91 center at 0 0
\endpicture$}
  \caption{The Circle of Octaves in Pyth-3}
  \label{figure-circle-of-octaves}
\end{figure}

We obtain:

\begin{enumerate}
\item Each note in Pyth-2 is octave-equivalent
      to a unique note in the circle of fifths.
\item Each note in Pyth-3 is tritave-equivalent 
      to a unique note in the circle of octaves.
\end{enumerate}

\subsection{Unique notation for each note in Pyth-3}

Next we introduce unique notation for each note in Pyth-3.
For comparison, recall that the 12 notes in Pyth-2 which lie in the fundamental domain
$I_2=(\kappa/\sqrt2,\kappa\,\sqrt2\,]$, sorted by scale degree, are denoted by

\smallskip
\centerline{\A,\, \B\f,\, \B,\, \C,\, \C\s,\, \D,\, \E\f,\, \E,\,
      \F,\, \F\s,\, \G,\, \G\s.}

\smallskip\noindent
An arbitrary note in Pyth-2 is octave-equivalent to one of the notes in the list;
if it is $p$ octaves higher than the note in the list, or $q$ octaves lower,
then we denote the note by adding $p$ primes or $q$ commas, respectively.
So, for example, \C\p\ is one octave above the \C\ and
\G\c\ one octave below the \G.

Similarly, the notes in the fundamental domain $I_3$, sorted by scale
degree, are

\smallskip
\centerline{\F\c,\, \F\s\c,\,  \G\c,\,   \A\f,\,  \A,\,  \B\f,\,  \B,\,  \C,\,
\C\s,\,  \D,\,  \E\f,\,  \E,\,  \F,\,  \F\s,\,  \G,\,  \G\s,\,
\A\p,\, \B\f\p,\, \B\p.}

\smallskip\noindent
Their up and down shifts by multiples of a tritave are
indicated by hats and check marks.

We emphasize that each note in Pyth-3 has a unique name.
This follows from the statement at the end of Section~\ref{subsection-circle-of-octaves}.

For example, \C\H\ and \G\V\ are the notes
one tritave above the \C\ and one tritave below the \G, respectively.
The notation \G\p\ and \C\c\ is reserved for Pyth-2;
the corresponding notes with the same frequencies are denoted in Pyth-3
by \C\H\ and \G\V, respectively.
Here is another example.  The note \F\s\c\H\ should perhaps be written as
(\F\s\c)\H\ to emphasize that it is obtained from the note \F\s\c\
(of harmonic degree -7 in the fundamental domain for Pyth-3)
by shifting it up by one tritave
(the expression \C\s\p\ is not used to denote any note in Pyth-3).
For each of the 88 notes on the piano keyboard, the
unique Pyth-3 label is listed
in Appendix~\ref{appendix-labels}.

\subsection{Playing Pyth-3 on the piano}
\label{subsection-piano}
Is it possible to play in a tritave-based system on the piano?

\smallskip
The answer is: yes, it's almost accurate --- and we will make is easy.

\smallskip\noindent
Before we start doing so, we need to caution the reader that our goal here
is only to show feasibility.  For a discussion of
measures with respect to which scales
are to be optimized, and for methods to accomplish this, we refer the reader
to \cite{polansky} and to \cite{krantz}.

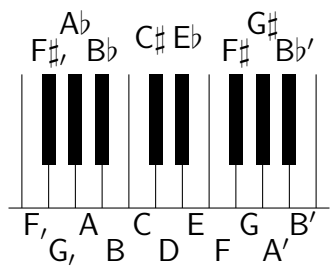
\begin{figure}[ht]
\begin{center}
\setlength\unitlength{.14in}
\begin{picture}(33,10)(0,-2)
\put(10.5,0){\line(1,0){12}}
\multiput(11,0)(1,0){12}{\line(0,1){5}}
\linethickness{.07in}
\put(11,0){\multiput(5,1.6)(1,0)2{\line(0,1){3.4}}}
\put(11,0){\multiput(1,1.6)(1,0)3{\line(0,1){3.4}}}
\put(11,0){\multiput(8,1.6)(1,0)3{\line(0,1){3.4}}}
\put(11.5,-1){\makebox[0ex]{\F\c}}
\put(12.5,-2){\makebox[0ex]{\G\c}}
\put(13.5,-1){\makebox[0ex]{\A}}
\put(14.5,-2){\makebox[0ex]{\B}}
\put(15.5,-1){\makebox[0ex]{\C}}
\put(16.5,-2){\makebox[0ex]{\D}}
\put(17.5,-1){\makebox[0ex]{\E}}
\put(18.5,-2){\makebox[0ex]{\F}}
\put(19.5,-1){\makebox[0ex]{\G}}
\put(20.5,-2){\makebox[0ex]{\A\p}}
\put(21.5,-1){\makebox[0ex]{\B\p}}
\put(12,5.5){\makebox[0ex]{\F\s\c}}
\put(13,6.5){\makebox[0ex]{\A\f}}
\put(14,5.5){\makebox[0ex]{\B\f}}
\put(15.8,6){\makebox[0ex]{\C\s}}
\put(17.2,6){\makebox[0ex]{\E\f}}
\put(19,5.5){\makebox[0ex]{\F\s}}
\put(20,6.5){\makebox[0ex]{\G\s}}
\put(21.2,5.5){\makebox[0ex]{\B\f\p}}
\end{picture}
\end{center}
  \caption{The Fundamental Domain for Pyth-3}
  \label{figure-fun19}
\end{figure}

In Figure~\ref{figure-fun19}, we list the notes from the circle of octaves,
they occur as keys on the piano (for the \A\f\ see below).
The remaining notes in Pyth-3 are obtained from those by going up
or down a certain number of tritaves
(see Appendix~\ref{appendix-labels}).

The point is that the notes in Pyth-2 and in Pyth-3 are almost the same.
Yet for composition and in terms of harmonic properties, octave-based
systems and tritave-based systems are formally treated in a different way.

Let us first consider the piano in just Pythagorean intonation.
We have seen in Section~\ref{subsection-enharmonic} that each note in
Pyth-2 is enharmonic to one in Pyth-3, and for most notes on the piano
they are equal.  They differ only for the eight notes
listed in Table~\ref{table-twonotthree},
and there, the difference is plus or minus one comma.

\begin{table}[ht]
\label{table-twonotthree}
\caption{Scale degrees where Pyth-2 and Pyth-3 differ}
{\begin{tabular}{|l||c|c|c|c|c|c|c|c|}
\hline
\multicolumn9{|c|}{{\bf Where Pyth-2 and Pyth-3 differ in just intonation}} \\
\hline \hline
Scale degree & -37 & -30 & -25 & -18 & -6 & 25 & 37 & 44 \\ \hline
Note in Pyth-3 &  \E\f\V\V\ & \B\f\p\V\V\ & \A\f\V\ &
\E\f\V\ & \A\f\ & \G\s\H\ & \C\s\H\H\ & \G\s\H\H\ \\ \hline
Note in Pyth-2 & \C\s\c\c\c\ & \G\s\c\c\c\ & \C\s\c\c\ & \G\s\c\c\ &
\G\s\c\ & \E\f\p\p\ & \E\f\p\p\p\ & \B\f\p\p\p\p\ \\ \hline
Frequency quotient & $\frac1\kappa$ & $\frac1\kappa$ & $\frac1\kappa$ &
$\frac1\kappa$ & $\frac1\kappa$ & $\kappa$ & $\kappa$ & $\kappa$ \\ \hline
\end{tabular}}
\end{table}

The first place where they differ is at scale degree $-6$, here the note in
Pyth-2 is the \G\s\c\ of frequency $2^{-10}\cdot 3^6$, while
Pyth-3 has at this position the \A\f\ of frequency
$2^9\cdot 3^{-6}=2^{-10}\cdot3^6/ \kappa$  (in Pyth-2,
we give preference to a note of harmonic degree $\mu_3=6$
over harmonic degree $\mu_3=-6$, while 
in Pyth-3, we prefer
harmonic degree $\mu_2=9$ over $\mu_2=-10$).
Table~\ref{table-twonotthree} lists all notes on the 88-key piano where
Pyth-2 and Pyth-3 differ.

Next we consider the piano in  equal temperament.  The note at scale degree $n$
is commonly
tuned to frequency $2^{\frac{n}{12}}$.  This system is called 12-EDO
(equal division of the octave).

For equal intonation in a system of 19 notes per tritave,
we simply assign to the note at scale degree $n$
the ratio $3^{\frac n{19}}$.  We call this system 19-EDT.
We have seen in Table~\ref{table-19edt} that each note in Pyth-3
in just intonation
differs by at most $\pm11.11$ cents from the corresponding note in 19-EDT.

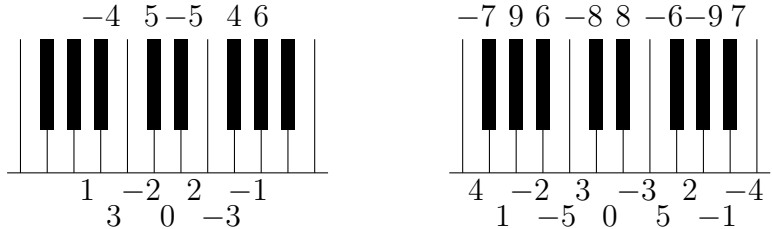
\begin{figure}[ht]
  \begin{center}
    \setlength\unitlength{.14in}
    \begin{picture}(12,8)(9.5,-2.5)
      \put(9.5,0){\line(1,0){12}}
      \multiput(10,0)(1,0){12}{\line(0,1){5}}
      \linethickness{.07in}
      \multiput(14,0)(7,0)1{\multiput(1,1.6)(1,0)2{\line(0,1){3.4}}}
      \multiput(7,0)(7,0)2{\multiput(4,1.6)(1,0)3{\line(0,1){3.4}}}
      \put(12.5,-1){\makebox[0ex]{$1$}}
      \put(13.5,-2){\makebox[0ex]{$3$}}
      \put(14.5,-1){\makebox[0ex]{$-2$}}
      \put(15.5,-2){\makebox[0ex]{$0$}}
      \put(16.5,-1){\makebox[0ex]{$2$}}
      \put(17.5,-2){\makebox[0ex]{$-3$}}
      \put(18.5,-1){\makebox[0ex]{$-1$}}
      \put(13,5.5){\makebox[0ex]{$-4$}}
      \put(14.9,5.5){\makebox[0ex]{$5$}}
      \put(16.1,5.5){\makebox[0ex]{$-5$}}
      \put(18,5.5){\makebox[0ex]{$4$}}
      \put(19,5.5){\makebox[0ex]{$6$}}
    \end{picture}
    \qquad\qquad
    \begin{picture}(11,10)(11,-2.5)
      \put(10.5,0){\line(1,0){12}}
      \multiput(11,0)(1,0){12}{\line(0,1){5}}
      \linethickness{.07in}
      \multiput(11,0)(11,0)1{\multiput(5,1.6)(1,0)2{\line(0,1){3.4}}}
      \multiput(11,0)(11,0)1{\multiput(1,1.6)(1,0)3{\line(0,1){3.4}}}
      \multiput(11,0)(11,0)1{\multiput(8,1.6)(1,0)3{\line(0,1){3.4}}}
      \put(11.5,-1){\makebox[0ex]{$4$}}
      \put(12.5,-2){\makebox[0ex]{$1$}}
      \put(13.5,-1){\makebox[0ex]{$-2$}}
      \put(14.5,-2){\makebox[0ex]{$-5$}}
      \put(15.5,-1){\makebox[0ex]{$3$}}
      \put(16.5,-2){\makebox[0ex]{$0$}}
      \put(17.5,-1){\makebox[0ex]{$-3$}}
      \put(18.5,-2){\makebox[0ex]{$5$}}
      \put(19.5,-1){\makebox[0ex]{$2$}}
      \put(20.5,-2){\makebox[0ex]{$-1$}}
      \put(21.5,-1){\makebox[0ex]{$-4$}}
      \put(11.5,5.5){\makebox[0ex]{$-7$}}
      \put(13,5.5){\makebox[0ex]{$9$}}
      \put(14,5.5){\makebox[0ex]{$6$}}
      \put(15.5,5.5){\makebox[0ex]{$-8$}}
      \put(17,5.5){\makebox[0ex]{$8$}}
      \put(18.5,5.5){\makebox[0ex]{$-6$}}
      \put(20,5.5){\makebox[0ex]{$-9$}}
      \put(21.3,5.5){\makebox[0ex]{$7$}}
    \end{picture}
    \caption{Keyboard with harmonic degrees:  Pyth-2 (left) and Pyth-3}
    \label{fig-diatonic-keyboard}
  \end{center}
\end{figure}

Let us compare 19-EDT with 12-EDO.
Note that the numbers $3^{\frac1{19}}$ and $2^{\frac1{12}}$ are very similar.
Since $$\left(\frac{3^{\frac1{19}}}{2^{\frac1{12}}}\right)^{12\cdot 19}=
\frac{3^{12}}{2^{19}}=\kappa,$$
the difference between $3^{\frac1{19}}$ and $2^{\frac1{12}}$ in cents
is just $(1/(12\cdot 19)=1/228)$ times ($\kappa$ in cents), that is,
around $23.46/228\approx 0.103$ cents.

We restrict this comparison to the notes on a piano with 88 keys.
If the middle
notes are in tune, then for any scale degree the note in 12-EDO
differs by less than $\pm5$ cents from the note in 19-EDT of the same scale
degree.

We have already seen that each note in 19-EDT differs by
at most $\pm11.11$ cents from its corresponding note in Pyth-3.

In this sense, both 19-EDT and Pyth-3 in just intonation can be called
``piano scales'', as in the title of this section.
In particular, we will be using the usual music notation for all scales.

For the convenience of the reader, we attach in Appendix~\ref{appendix-labels}
key labels for Pyth-3 for an 88-key piano keyboard to exhibit
the notes with hats and check marks.

\section{Visualizing harmony:  the {\it Tonnetz}}

First, we briefly revisit the {\it Tonnetz} for 4:5:6 harmony, then adapt it to our situation.

\subsection{4:5:6 harmony}

The {\it Tonnetz} (German: tone-network) has been first described by Leonhard Euler,
see in particular \cite{euler}, and then was used in 19th century
musicology as the space in which harmonic development and key changes can be traced.
It plays an important role in neo-Riemannian theory, see Section~\ref{section-neo-Riemannian}.

\begin{figure}[ht]
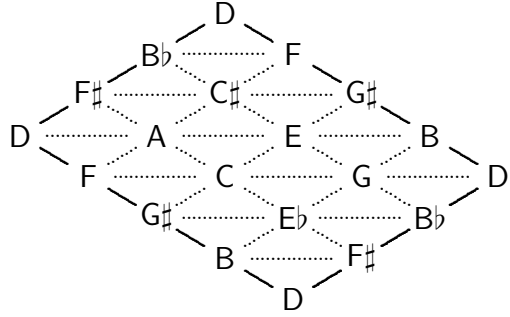

    \centerline{$
    \beginpicture
    \setcoordinatesystem units <5ex,3ex>
    \multiput{} at -7 -1  0 2 /
    \put {\D} at -7 1
    \put {\F\s} at -6 2
    \put {\F} at -6 0
    \put {\B\f} at -5 3
    \put {\A} at -5 1
    \put {\G\s} at -5 -1
    \put {\D} at -4 4
    \put {\C\s} at -4 2
    \put {\C} at -4 0
    \put {\B} at -4 -2
    \put {\F} at -3 3
    \put {\E} at -3 1
    \put {\E\f} at -3 -1
    \put {\D} at -3 -3
    \put {\G\s} at -2 2
    \put {\G} at -2 0
    \put {\F\s} at -2 -2
    \put {\B} at -1 1
    \put {\B\f} at -1 -1
    \put {\D} at 0 0
    \setsolid
    \plot -6.7 1.3  -6.3 1.7 /
    \plot -5.7 2.3  -5.3 2.7 /
    \plot -4.7 3.3  -4.3 3.7 /
    \plot -3.7 3.7  -3.3 3.3 /
    \plot -2.7 2.7  -2.3 2.3 /
    \plot -1.7 1.7  -1.3 1.3 /
    \plot -.7 .7    -.3 .3 /
    \plot -.3 -.3  -.7 -.7 /
    \plot -1.3 -1.3  -1.7 -1.7 /
    \plot -2.3 -2.3  -2.7 -2.7 /
    \plot -3.3 -2.7  -3.7 -2.3 /
    \plot -4.3 -1.7  -4.7 -1.3 /
    \plot -5.3 -.7   -5.7 -.3 /
    \plot -6.3 .3    -6.7 .7 /
    \setdots<2pt>
    \plot -5.7 .3  -5.3 .7 /
    \plot -4.7 -.7  -4.3 -.3 /
    \plot -3.7 -1.7  -3.3 -1.3 /
    \plot -2.7 -2.7  -2.3 -2.3 /
    \plot -4.7 1.3  -4.3 1.7 /
    \plot -3.7 .3  -3.3 .7 /
    \plot -2.7 -.7  -2.3 -.3 /
    \plot -1.7 -1.7  -1.3 -1.3 /
    \plot -3.7 2.3  -3.3 2.7 /
    \plot -2.7 1.3  -2.3 1.7 /
    \plot -1.7 .3  -1.3  .7 /
    \plot -.7 -.7  -.3  -.3 /
    \plot -5.7 1.7  -5.3 1.3 /
    \plot -4.7 .7  -4.3 .3 /
    \plot -3.7 -.3  -3.3 -.7 /
    \plot -2.7 -1.3  -2.3 -1.7 /
    \plot -4.7 2.7 -4.3 2.3 /
    \plot -3.7 1.7 -3.3 1.3 /
    \plot -2.7 .7  -2.3 .3 /
    \plot -1.7 -.3  -1.3 -.7 /
    \plot -4.6 3  -3.4 3 /
    \plot -5.6 2  -4.4 2 /
    \plot -3.6 2  -2.4 2 /
    \plot -6.6 1  -5.4 1 /
    \plot -4.6 1  -3.4 1 /
    \plot -2.6 1  -1.4 1 /
    \plot -5.6 0  -4.4 0 /
    \plot -3.6 0  -2.4 0 /
    \plot -1.6 0  -.4 0 /
    \plot -4.6 -1  -3.4 -1 /
    \plot -2.6 -1  -1.4 -1 /
    \plot -3.6 -2  -2.4 -2 /
    \endpicture$}
    \caption{{\it Tonnetz} for 4:5:6 harmony (one fundamental tile)}
    \label{figure-diatonic-fundamental}
  \end{figure}

  The original {\it Tonnetz} is infinite in each direction in the plane.
  In horizontal direction, the notes progress as given by the circle
  of fifths, corresponding to the ratio 6:4 or 3:2.
  The horizontal step is divided into two parts given by the two small
  intervals in the chord:  the major third (5:4), pictured as an upwards step,
  and the minor third (6:5), represented by a downwards step.
  We picture in Figure~\ref{figure-diatonic-fundamental}
  one fundamental tile.  The {\it infinite Tonnetz} can be obtained
  by connecting  different copies of the fundamental tile along   parallel solid lines.
  The note labels need to be adjusted to reflect that there are perfect fifths in
  horizontal direction (so the note on the left of the \E\f\ is to be labelled \A\f, not \G\s)
  and major thirds along the upwards diagonals.

In neo-Riemannian theory it is common to use equal intonation (12-EDO) and to identify
octave-equivalent notes.  This yields the {\it finite Tonnetz} which, topologically, is best represented as
the torus obtained by identifying parallel solid lines in the fundamental tile in
Figure~\ref{figure-diatonic-fundamental}.

\subsection{The \boldit{Tonnetz} for 2:3:4 harmony}

\begin{figure}[ht]
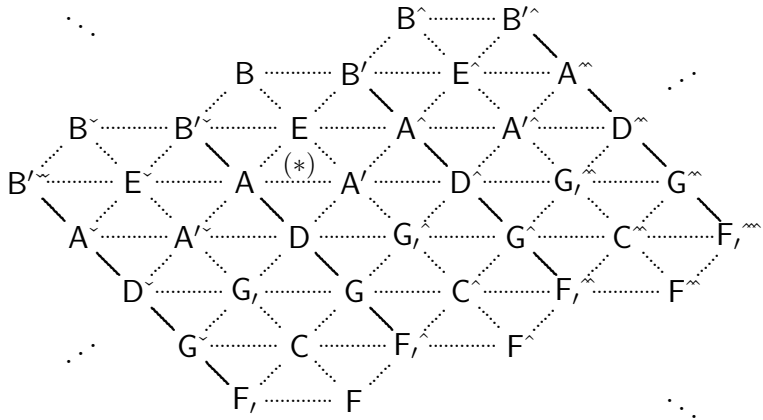

    \centerline{$
    \beginpicture
    \setcoordinatesystem units <4ex,4ex>
    \multiput{} at -2 5  10 -1 /
    \put {\D\V} at 0 0 
    \put {\A\V} at -1 1
    \put {\B\p\V\V} at -2 2
    \put {\G\V} at 1 -1 
    \put {\F\c} at 2 -2
    \put {\F} at 4 -2
    \put {\C} at 3 -1
    \put {\G\c} at 2 0 
    \put {\A\p\V} at 1 1 
    \put {\E\V} at 0 2
    \put {\B\V} at -1 3 
    \put {\D} at 3 1 
    \put {\A} at   2 2
    \put {\B\p\V} at   1 3 
    \put {\G} at  4 0 
    \put {\F\c\H} at  5 -1 
    \put {\F\H} at 7 -1
    \put {\C\H} at 6 0
    \put {\G\c\H} at 5 1 
    \put {\A\p} at 4 2 
    \put {\E} at 3 3
    \put {\B} at 2 4 
    \put {\D\H} at 6 2 
    \put {\A\H} at   5 3
    \put {\B\p} at   4 4 
    \put {\G\H} at  7 1 
    \put {\F\c\H\H} at  8 0
    \put {\F\H\H} at 10 0
    \put {\C\H\H} at 9 1
     \put {\A\H\H} at 8 4
     \put {\B\p\H} at 7 5
    \put {\G\c\H\H} at 8 2
    \put {\A\p\H} at 7 3 
    \put {\E\H} at 6 4
    \put {\B\H} at 5 5
    \put {\D\H\H} at 9 3
    \put {\G\H\H} at 10 2
    \put {\F\c\H\H\H} at 11 1
    \put {\footnotesize $(*)$} at 3 2.3
    \multiput {\reflectbox{$\ddots$}} at 10 4  -1 -1  /
    \multiput {$\ddots$} at 10 -2  -1 5 /
    \setsolid
    \plot 1.7 -1.7  1.3 -1.3 /
    \plot .7 -.7  .3 -.3 /
    \plot -.3 .3  -.7 .7 /
    \plot -1.3 1.3  -1.7 1.7 /
    \plot 4.7 -.7  4.3 -.3 /
    \plot 3.7 .3  3.3 .7 /
    \plot 2.7 1.3  2.3 1.7 /
    \plot 1.7 2.3  1.3 2.7 /
    \plot 7.7 .3  7.3 .7 /
    \plot 6.7 1.3  6.3 1.7 /
    \plot 5.7 2.3  5.3 2.7 /
    \plot 4.7 3.3  4.3 3.7 /
    %
     \plot 7.3 4.7  7.7 4.3 /
     \plot 8.3 3.7  8.7 3.3 /
     \plot 9.3 2.7  9.7 2.3 /
     \plot 10.3 1.7  10.7 1.3 /
     \setdots<2pt>
     \plot -1.7 2.3  -1.3 2.7 /
     \plot -.7 1.3  -.3 1.7 /
     \plot .3 .3  .7 .7 /
     \plot 1.3 -.7  1.7 -.3 /
     \plot 2.3 -1.7  2.7 -1.3 /
     \plot .3 2.3  .7 2.7 /
     \plot 1.3 1.3  1.7 1.7 /
     \plot 2.3 .3  2.7 .7 /
     \plot 3.3 -.7  3.7 -.3 /
     \plot 4.3 -1.7  4.7 -1.3 /
     \plot 1.3 3.3  1.7 3.7 /
     \plot 2.3 2.3  2.7 2.7 /
     \plot 3.3 1.3  3.7 1.7 /
     \plot 4.3 .3   4.7 .7 /
     \plot 5.3 -.7  5.7 -.3 /
     \plot 3.3 3.3  3.7 3.7 /
     \plot 4.3 2.3  4.7 2.7 /
     \plot 5.3 1.3  5.7 1.7 /
     \plot 6.3 .3   6.7 .7 /
     \plot 7.3 -.7  7.7 -.3 /
     \plot 4.3 4.3  4.7 4.7 /
     \plot 5.3 3.3  5.7 3.7 /
     \plot 6.3 2.3  6.7 2.7 /
     \plot 7.3 1.3  7.7 1.7 /
     \plot 8.3 .3   8.7 .7 /
     \plot 6.3 4.3  6.7 4.7 /
     \plot 7.3 3.3  7.7 3.7 /
     \plot 8.3 2.3  8.7 2.7 /
     \plot 9.3 1.3  9.7 1.7 /
     \plot 10.3 .3  10.7 .7 /
     \plot -.7 2.7  -.3 2.3 /
     \plot .3 1.7  .7 1.3 /
     \plot 1.3 .7  1.7 .3 /
     \plot 2.3 -.3  2.7 -.7 /
     \plot 3.3 -1.3  3.7 -1.7 /
     \plot 2.3 3.7  2.7 3.3 /
     \plot 3.3 2.7  3.7 2.3 /
     \plot 4.3 1.7  4.7 1.3 /
     \plot 5.3 .7   5.7 .3 /
     \plot 6.3 -.3  6.7 -.7 /
     \plot 5.3 4.7  5.7 4.3 /
     \plot 6.3 3.7  6.7 3.3 /
     \plot 7.3 2.7  7.7 2.3 /
     \plot 8.3 1.7   8.7 1.3 /
     \plot 9.3 .7   9.7 .3 /
	\plot 5.4 5  6.6 5 /
	\plot 2.4 4  3.6 4 /
	\plot 4.4 4  5.6 4 /
	\plot 6.4 4  7.6 4 /
	\plot -.6 3  .6 3 /
	\plot 1.4 3  2.6 3 /
	\plot 3.4 3  4.6 3 /
	\plot 5.4 3  6.6 3 /
	\plot 7.4 3  8.6 3 /
        \plot -1.6 2  -.4 2 /
	\plot .4 2  1.6 2 /
	\plot 2.4 2  3.6 2 /
	\plot 4.4 2  5.6 2 /
	\plot 6.4 2  7.6 2 /
	\plot 8.4 2  9.6 2 /
        \plot -.6 1  .6 1 /
	\plot 1.4 1  2.6 1 /
	\plot 3.4 1  4.6 1 /
	\plot 5.4 1  6.6 1 /
	\plot 7.4 1  8.6 1 /
	\plot 9.4 1  10.6 1 /
	\plot .4 0  1.6 0 /
	\plot 2.4 0  3.6 0 /
	\plot 4.4 0  5.6 0 /
	\plot 6.4 0  7.6 0 /
	\plot 8.4 0  9.6 0 /
	\plot 1.4 -1  2.6 -1 /
	\plot 3.4 -1  4.6 -1 /
	\plot 5.4 -1  6.6 -1 /
	\plot 2.4 -2  3.6 -2 /
    \endpicture$}
    \caption{Part of the infinite 2:3:4 {\it Tonnetz}}
    \label{figure-19notes-extended}
  \end{figure}

We introduce the {\it Tonnetz for 2:3:4 harmony,} more precisely,
the finite 2:3:4 {\it Tonnetz} and the
infinite 2:3:4 {\it Tonnetz.}
In horizontal direction, the notes progress by octaves (ratio 4:2 or 2:1),
each step is divided into an upwards move given by the fifth (3:2)
and a downwards move given by the fourth (4:3).
In Figure~\ref{figure-19notes-extended}, we picture the notes around the
\A-\E-\A\p\ chord which we label by an (*).

\begin{figure}[ht]
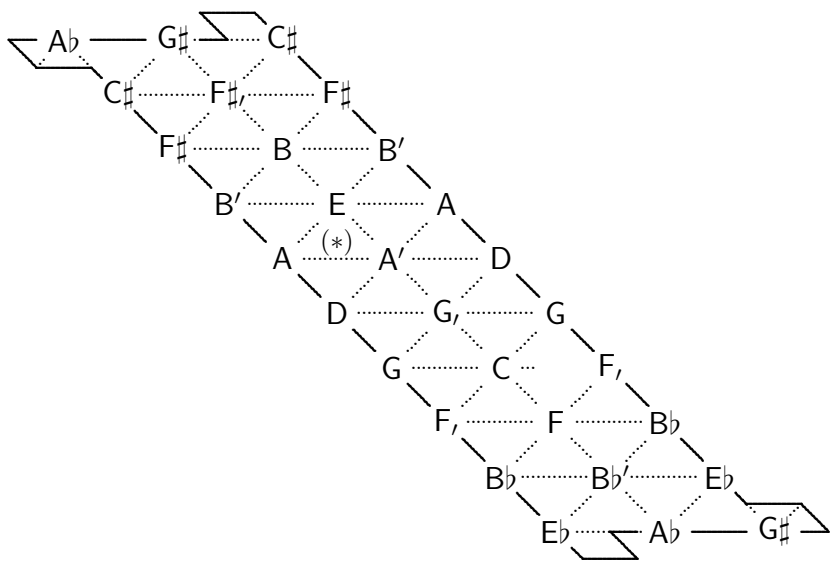

    \centerline{$
    \beginpicture
    \setcoordinatesystem units <4ex,4ex>
    \multiput{} at -6 5.5  9 -4.5  /
    \multiput {\D} at 0 0  3 1 /
    \multiput {\A} at -1 1  2 2 /
    \multiput {\B\p} at -2 2  1 3 /
    \multiput {\F\s} at -3 3  0 4 /
    \multiput {\C\s} at -4 4  -1 5 /
    \multiput {\A\f} at -5 5  6 -4 /
    \multiput {\G\s} at -3 5  8 -4 /
    \multiput {\G} at 1 -1  4 0 /
    \multiput {\F\c} at 2 -2  5 -1 /
    \multiput {\B\f} at 3 -3  6 -2 /
    \multiput {\E\f} at 4 -4  7 -3 /
    \put {\B\f\p} at 5 -3 
    \put {\F} at 4 -2
    \put {\C} at 3 -1
    \put {\G\c} at 2 0 
    \put {\A\p} at 1 1 
    \put {\E} at 0 2
    \put {\B} at -1 3 
    \put {\F\s\c} at -2 4 
    \put {\footnotesize $(*)$} at 0 1.3
    \setsolid
    \plot 6.5 -4  7.5 -4 /  
    \plot 5.5 -4  5 -4  5.5 -4.5  4.5 -4.5  4.3 -4.3 /
    \plot 3.7 -3.7  3.3 -3.3 /
    \plot 2.7 -2.7  2.3 -2.3 /
    \plot 1.7 -1.7  1.3 -1.3 /
    \plot .7 -.7  .3 -.3 /
    \plot -.3 .3  -.7 .7 /
    \plot -1.3 1.3  -1.7 1.7 /
    \plot 8.5 -4  9 -4  8.5 -3.5  7.5 -3.5  7.3 -3.3 /
    \plot 6.7 -2.7  6.3 -2.3 /
    \plot 5.7 -1.7  5.3 -1.3 /
    \plot 4.7 -.7  4.3 -.3 /
    \plot 3.7 .3  3.3 .7 /
    \plot 2.7 1.3  2.3 1.7 /
    \plot 1.7 2.3  1.3 2.7 /
    \plot .7 3.3  .3 3.7 /
    \plot -.3 4.3  -.7 4.7 /
    \plot -1.3 5.3  -1.5 5.5  -2.5 5.5  -2 5  -2.5 5 /
    \plot -3.5 5  -4.5 5 /
    \plot -5.5 5  -6 5  -5.5 4.5  -4.5 4.5  -4.3 4.3  /
    \plot -3.7 3.7  -3.3 3.3 /
    \plot -2.7 2.7  -2.3 2.3 /
    \setdots<2pt>
    \plot -4.7 4.7  -4.3 4.3 /
    \plot -2.7 4.7  -2.3 4.3 /
    \plot -1.7 3.7  -1.3 3.3 /
    \plot -.7 2.7  -.3 2.3 /
    \plot .3 1.7  .7 1.3 /
    \plot 1.3 .7  1.7 .3 /
    \plot 2.3 -.3  2.7 -.7 /
    \plot 3.3 -1.3  3.7 -1.7 /
    \plot 4.3 -2.3  4.7 -2.7 /
    \plot 5.3 -3.3  5.7 -3.7 /
    \plot 7.3 -3.3  7.7 -3.7 /
    \plot -5.5 4.5  -5.3 4.7 /
    \plot -3.7 4.3  -3.3 4.7 /
    \plot -1.7 4.3  -1.3 4.7 /
    \plot -2.7 3.3  -2.3 3.7 /
    \plot -.7  3.3  -.3  3.7 /
    \plot -1.7 2.3  -1.3 2.7 /
    \plot .3 2.3  .7 2.7 /
    \plot -.7 1.3  -.3 1.7 /
    \plot 1.3 1.3  1.7 1.7 /
    \plot .3 .3  .7 .7 /
    \plot 2.3 .3  2.7 .7 /
    \plot 1.3 -.7  1.7 -.3 /
    \plot 3.3 -.7  3.7 -.3 /
    \plot 2.3 -1.7  2.7 -1.3 /
    \plot 4.3 -1.7  4.7 -1.3 /
    \plot 3.3 -2.7  3.7 -2.3 /
    \plot 5.3 -2.7  5.7 -2.3 /
    \plot 4.3 -3.7  4.7 -3.3 /
    \plot 6.3 -3.7  6.7 -3.3 /
    \plot 8.3 -3.7  8.5 -3.5 /
    \plot -2.6 5  -1.4 5 /
    \plot -3.6 4  -2.4 4 /
    \plot -1.6 4  -.4 4 /
    \plot -2.6 3  -1.4 3 /
    \plot -.6 3  .6 3 /
    \plot -1.6 2  -.4 2 /
    \plot .4 2  1.6 2 /
    \plot -.6 1  .6 1 /
    \plot 1.4 1  2.6 1 /
    \plot .4 0  1.6 0 /
    \plot 2.4 0  3.6 0 /
    \plot 1.4 -1  2.6 -1 /
    \plot 3.4 -1  3.6 -1 /
    \plot 2.4 -2  3.6 -2 /
    \plot 4.4 -2  5.6 -2 /
    \plot 3.4 -3  4.6 -3 /
    \plot 5.4 -3  6.6 -3 /
    \plot 4.4 -4  5.6 -4 /
    \endpicture$}
    \caption{{\it Tonnetz} for 2:3:4 harmony (one fundamental tile)}
    \label{figure-19notes-full}
  \end{figure}

The {\it infinite 2:3:4 Tonnetz} is obtained by placing copies of the fundamental tile in
Figure~\ref{figure-19notes-full} next to each other, as indicated by the solid lines.
Hooks are used to indicate how the solid lines are to be shifted against each other.
In horizontal direction, hats and checkmarks need to be added so that in each step, the frequency
increases by an octave, as in Figure~\ref{figure-19notes-extended}.
In vertical direction, the notes and their labels need to be adjusted by Pythagorean commas;
for example,
the top left note in the fundamental tile should be a \G\s\c, not an \A\f, and the bottom right note
an \A\f\p, not a \G\s.

For the {\it finite 2:3:4 Tonnetz,} we use equal intonation (19-EDT) and identify tritave-equivalent
notes.  Topologically, the {\it Tonnetz} is best represented as the
torus obtained by identifying opposite sides
of the fundamental tile in Figure~\ref{figure-19notes-full}.

\empar{Major and minor chords}
One can define {\it major} and {\it minor} chords for 2:3:4 harmony as chords where the fifth
is followed by a fourth, or the fourth followed by a fifth, respectively.
So the 2:3:4 chord \A-\E-\A\p\ is a major chord, and the 3:4:6 chord \A-\D-\A\p\
a minor chord.
Similarly, an {\it augmented} or {\it diminished} chord consists of two fifths
or two fourths, respectively.

There is an important difference for major and minor 2:3:4 chords, when compared
with 4:5:6 harmony:  the first inversion of a major chord turns out to be
a minor chord.  For example, the first inversion of \A-\E-\A\p\ is the minor
chord \E-\A\p-\A\H, pictured on the right hand side of (*) in the {\it Tonnetz.}

We will discuss this further in Section~\ref{section-inversions}.

\subsection{Neo-Riemannian theory for 2:3:4 chords}
\label{section-neo-Riemannian}

In neo-Riemannian theory, chord progressions are analyzed in terms of elementary moves
or {\it transformations} in the {\it Tonnetz.}
For example, the process of going from tonic to dominant (which differs from the tonic chord
in two notes) is decomposed as a product of two elementary moves of which each changes
only one note.

The three elementary moves considered are the P-, L-, and R-transformations;
they map a major or minor triad to the minor or major triad adjacent to one of the
edges of the triangle representing the chord in the {\it Tonnetz.}  See \cite{cohn}.

In this section we describe the three transformations when adapted to 2:3:4 harmony.
Here we need to work in the infinite 2:3:4 {\it Tonnetz} since major and minor triads are not
invariant under first and second inversions modulo the tritave.

    \begin{figure}[ht]
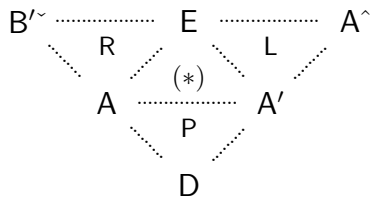

    \centerline{$
    \beginpicture
    \setcoordinatesystem units <6ex,6ex>
     \put {\D} at 3 1 
    \put {\A} at   2 2
    \put {\B\p\V} at   1 3 
     \put {\A\p} at 4 2 
    \put {\E} at 3 3
     \put {\A\H} at   5 3
     \setdots<2pt>
     \plot 1.4 3  2.6 3 /
     \plot 3.4 3  4.6 3 /
     \plot 2.4 2  3.6 2 /
     \plot 1.3 2.7  1.7  2.3 /
     \plot 2.3 1.7  2.7 1.3 /
     \plot 3.3 2.7  3.7 2.3 /
     \plot 2.3 2.3  2.7 2.7 /
     \plot 3.3 1.3  3.7 1.7 /
     \plot 4.3 2.3  4.7 2.7 /
     \put {\footnotesize $(*)$} at 3 2.3
     \put {\footnotesize\sf R} at 2 2.7
     \put {\footnotesize\sf L} at 4 2.7
     \put {\footnotesize\sf P} at 3 1.7
    \endpicture
    $}
    \caption{PLR-moves in the infinite 2:3:4 {\it Tonnetz}}
    \label{figure-PLR-1}
  \end{figure}

The P-transformation (parallel major or minor) maps a major triad to the minor triad with which it
has the octave in common, and conversely.  In the example, the major \A-\E-\A\p\ chord and the
minor \A-\D-\A\p\ chord correspond to each other under P-moves, see Figure~\ref{figure-PLR-1}.

Note that \A-\D-\A\p\ is the dominant chord for \A-\E-\A\p\, so a single move in 2:3:4 harmony
suffices to go from tonic to dominant, while two moves (R and L) are needed in 4:5:6 harmony.
Also note that the middle notes of the two chords differ by two semitones, while in
the 4:5:6 system, a P-transformation moves the middle note up or down by one semitone.

Under the R-transformation (relative major or minor), a major and a minor chord correspond to
each other whenever they have
the fifth in common.  In the example, an R-move maps the \A-\E-\A\p\ chord
to \B\p\V-\A-\E.

Note that the R-transformation for 
(*) is the second inversion of the subdominant chord \A-\E-\B\p.
In 2:3:4 harmony, it moves the last note
of the major chord up by two semitones, up to a tritave; this is similar to 4:5:6 harmony.

Finally, the L-transformation (lead tone exchange) maps a given 2:3:4 major or minor chord
to the one with which it shares the fourth.
This results in an inversion of the chord.
By comparison, in 4:5:6 harmony,
the lead tone of a major chord is replaced by a note one semitone lower.

\empar{Voice-leading parsimony}
An important application of neo-Riemannian theory is to voice leading
since the PLR-moves provide a measure for harmonic distance between chords.

In this paragraph we suppose that the harmony is given
by a sequence of triads which are to support a melody.  We ask: which notes may possibly
occur in the melody provided that
changes in the harmony are limited to a small number of PLR-moves?

We compare how many notes can be reached using a given number of PLR-moves in 4:5:6 harmony
with the corresponding numbers for 2:3:4 harmony. 
Table~\ref{tablePLR12} shows that in 4:5:6 harmony, with one exception (the \F\s\ if \C-\E-\G\ is the tonic),
any note is part of a triad
that can be reached from the tonic in at most two moves.

\begin{table}[ht]
  \small
\label{tablePLR12}
\caption{Notes in the 4:5:6 system reachable  by PLR-moves}
{\begin{tabular}{|r|cccc|}
 \hline
 number of transformations & $0$  & $1$ & $2$ & $3$ \\ \hline
 notes reachable up to octave-equivalence & $3$  & $6$  & $11$  & $12$  \\ \hline
\end{tabular}}
\end{table}

\begin{figure}[ht]
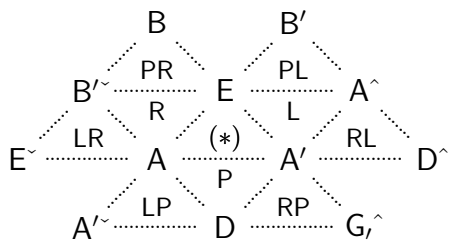

    \centerline{$    \beginpicture
    \setcoordinatesystem units <5ex,5ex>
    \put {\A\p\V} at 1 1 
    \put {\E\V} at 0 2
    \put {\D} at 3 1 
    \put {\A} at   2 2
    \put {\B\p\V} at   1 3 
    \put {\G\c\H} at 5 1 
    \put {\A\p} at 4 2 
    \put {\E} at 3 3
    \put {\B} at 2 4 
    \put {\D\H} at 6 2 
    \put {\A\H} at   5 3
    \put {\B\p} at   4 4 
    \put {\footnotesize $(*)$} at 3 2.3
    \setdots<2pt>
    \plot 0.4 2  1.6 2 /
    \plot 4.4 2  5.6 2 /
    \plot 1.4 1  2.6 1 /
    \plot 3.4 1  4.6 1 /
    \plot 1.4 3  2.6 3 /
     \plot 3.4 3  4.6 3 /
     \plot 2.4 2  3.6 2 /
     \plot 2.3 3.7  2.7 3.3 /
     \plot 4.3 1.7  4.7 1.3 /
     \plot 4.3 3.7  4.7 3.3 /
     \plot 5.3 2.7  5.7 2.3 /
     \plot 1.3 2.7  1.7  2.3 /
     \plot 2.3 1.7  2.7 1.3 /
     \plot 3.3 2.7  3.7 2.3 /
     \plot 0.3 2.3  0.7 2.7 /
     \plot 1.3 3.3  1.7 3.7 /
     \plot 1.3 1.3  1.7 1.7 /
     \plot 3.3 3.3  3.7 3.7 /
     \plot 2.3 2.3  2.7 2.7 /
     \plot 3.3 1.3  3.7 1.7 /
     \plot 4.3 2.3  4.7 2.7 /
     \put {\footnotesize $(*)$} at 3 2.3
     \put {\footnotesize\sf R} at 2 2.7
     \put {\footnotesize\sf L} at 4 2.7
     \put {\footnotesize\sf P} at 3 1.7
     \put {\footnotesize\sf LR} at 1 2.3
     \put {\footnotesize\sf PR} at 2 3.3
     \put {\footnotesize\sf LP} at 2 1.3
     \put {\footnotesize\sf PL} at 4 3.3
     \put {\footnotesize\sf RP} at 4 1.3
     \put {\footnotesize\sf RL} at 5 2.3
    \endpicture
    $}
    \caption{Notes in the 2:3:4 system reachable with up to two PLR-moves}
    \label{figure-PLR-2}
  \end{figure}

By comparison, in 2:3:4 harmony, only seven notes of 19 can be reached with up to two moves,
up to tritave-equivalence, see Table~\ref{tablePLR19}.
The notes which can be reached from the \A-\E-\A\p\ chord
and the corresponding PLR-moves are pictured in Figure~\ref{figure-PLR-2}.

\begin{table}[ht]
  \small
\label{tablePLR19}
\caption{Notes in the 2:3:4 system that can be reached by PLR-moves}
{\begin{tabular}{|r|ccccccccc|}
\hline
 number of transformations & $0$  & $1$ & $2$  & $3$
                   & $4$ & $5$ & $6$ & $7$ & $8$ \\ \hline
 notes reachable up to tritave-equiv.\ & $3$  & $5$  & $7$  & $9$ & $11$ & $13$ & $15$ & $17$ & $19$  \\
 \hline
\end{tabular}}
\end{table}

In summary, the PLR-moves of neo-Riemannian theory can be  adapted to 2:3:4 harmony.
They mimic closely composition elements related to dominants and subdominants,
inversions, and major and minor chords.
When compared to 4:5:6 harmony, fewer notes can be reached with a given number of PLR-moves.

\section{Elements of composition for 2:3:4 harmony}

We study basic elements of composition, which we adapt
to a tonal system that  is modulated with respect to the tritave.
We discuss perception of 2:3:4 harmony in the concluding Section~\ref{subsection-purity}
on purity and sparsity.

\subsection{Tonic, dominant and subdominant}
\label{subsection-dominant}

\begin{figure}[ht]
\label{figure-basic-sequence-456}
\begin{center}
\includegraphics[width=\linewidth]{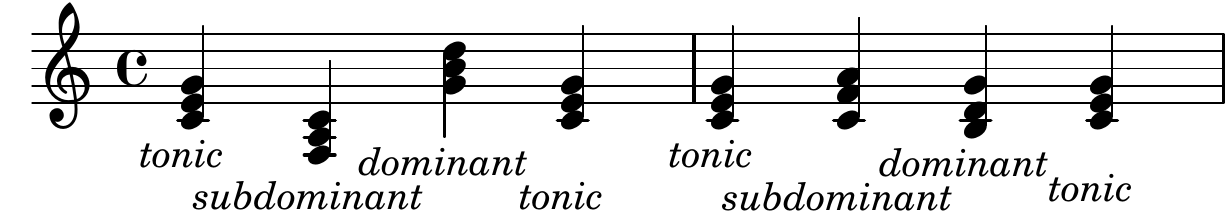}
\end{center}
\caption{The basic sequence in 4:5:6 harmony}
\end{figure}

We revisit the sequence tonic to subdominant to dominant to tonic
in 4:5:6 harmony.  Subdominant and dominant of a triad are obtained
by replacing each note by its predecessor or successor, respectively, in the cycle of fifths.
Starting from the major chord  \C-\E-\G, we obtain in Figure~\ref{figure-basic-sequence-456}
the following sequence,
which we repeat in the second bar using first and second
inversions of the triads modulo the octave.

\begin{figure}[ht]
\label{figure-basic-sequence-234}
\begin{center}
\includegraphics[width=\linewidth]{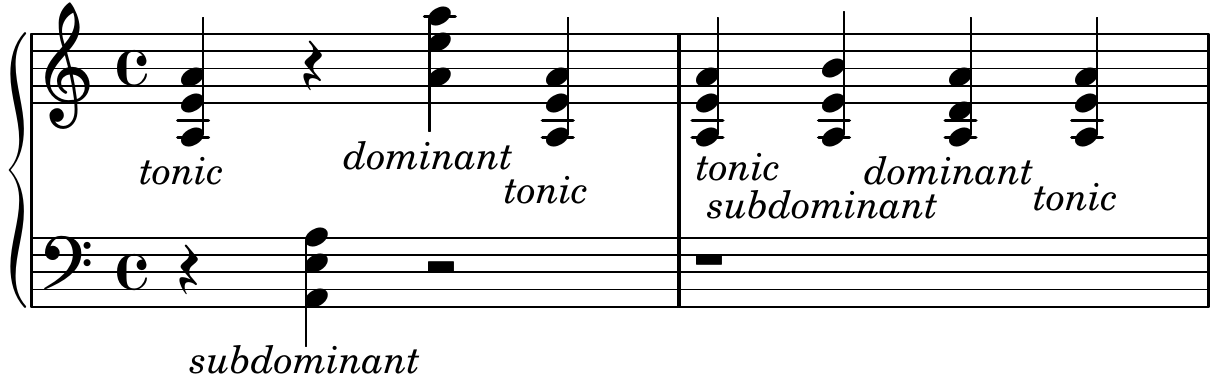}
\end{center}
\caption{The basic sequence in 2:3:4 harmony}
\end{figure}

We recreate this sequence in the 2:3:4 system.  Here, subdominant and dominant of a chord
are given by substituting each note by its predecessor or successor, respectively, in the cycle
of octaves.  Taking as tonic the \A-\E-\A\p\ chord, we obtain in Figure~\ref{figure-basic-sequence-234}
the following sequence.
In the second bar, we use inversions modulo the tritave to bring the triads into one fundamental domain.

\begin{figure}[ht]
\label{figure-finality-234}
\begin{center}
  \includegraphics{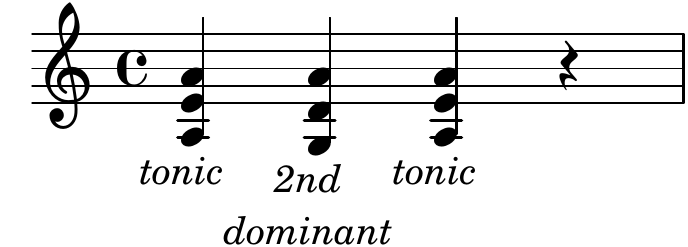}
\end{center}
\caption{Creating a sense of finality in 2:3:4 harmony}
\end{figure}

\empar{Cadences in 2:3:4 harmony}
It appears to the author that the step from dominant to tonic in
the above basic sequence in 2:3:4 harmony can be perceived as
a cadence, much like the corresponding step in 4:5:6 harmony.
A perhaps stronger sense of finality can be obtained by
using the step from the second dominant back to tonic,
as in the example in Figure~\ref{figure-finality-234}.

\empar{Dominant and subdominant in the \boldit{Tonnetz}}
The basic sequence in 4:5:6 harmony

\smallskip
\centerline{\C-\E-\G\ (\ding{192}) --- \C-\F-\A\ (\ding{193}) ---
		      \B-\D-\G\ (\ding{194}) --- \C-\E-\G\ (\ding{195})}

appears as follows in the {\it Tonnetz} in Figure~\ref{figure-basic-456}:

\begin{figure}[ht]
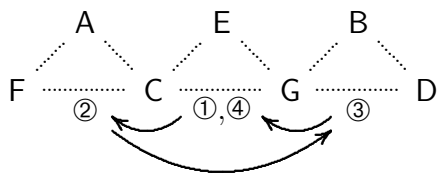

\centerline{$
    \beginpicture
    \setcoordinatesystem units <5ex,5ex>
    \put {\F} at -6 2
    \put {\A} at -5 3
    \put {\C} at -4 2
    \put {\E} at -3 3
    \put {\G} at -2 2
    \put {\B} at -1 3
    \put {\D} at 0 2
    \setdots<2pt>
    \put {\Ntri}  at -3 2
    \put {\ding{192},\ding{195}} at -3 1.7
    \put {\Ntri} at -1 2
    \put {\ding{194}} at -1 1.7
    \put {\Ntri} at -5 2
    \put {\ding{193}} at -5 1.7
    \setsolid
    \setquadratic
    \plot -3.6 1.6  -4.1 1.4  -4.6 1.6 /
    \arr{-4.5 1.53}{-4.6 1.6}
    \plot -1.4 1.6  -1.9 1.4  -2.4 1.6 /
    \arr{-2.3 1.53}{-2.4 1.6}
    \plot -4.6 1.4  -3 .8  -1.4 1.4 /
    \arr{-1.5 1.33}{-1.4 1.4}
    \endpicture
    $}
    \caption{The basic sequence in 4:5:6 harmony}
    \label{figure-basic-456}
    \end{figure}
    
Note that the 4:5:6 {\it Tonnetz} does not distinguish octave-equivalent, but different, notes.
Hence the notes in the  \C-\F-\A-triad (\ding{193}) appear as \F-\A-\C.

Here is the basic sequence for 2:3:4 harmony, before applying tritave-equivalence:

\smallskip
\centerline{\A-\E-\A\p\ (\ding{202}) --- \E\V-\B\p\V-\A\ (\ding{203}) ---
		      \A\p-\A\H-\D\H\ (\ding{204})
		      --- \A-\E-\A\p\ (\ding{205})}

\smallskip\noindent
We picture the sequence in the infinite 2:3:4 {\it Tonnetz.}
Note that the pitch increases in the direction of the arc.

\begin{figure}[ht]
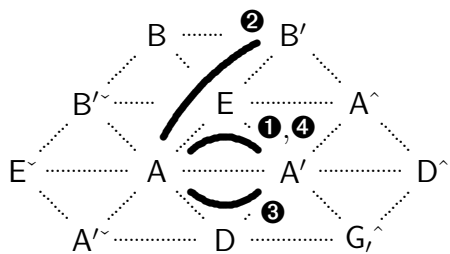

\centerline{$\beginpicture
    \setcoordinatesystem units <5ex,5ex>
    \put {\E\V} at 1 2
    \put {\A\p\V} at 2 1
    \put {\D} at 4 1
    \put {\G\c\H} at 6 1
    \put {\A} at  3 2
    \put {\A\p} at 5 2
    \put {\B\p\V} at 2 3
    \put {\E} at 4 3
    \put {\A\H} at 6 3 
    \put {\B} at 3 4 
    \put {\B\p} at  5 4 
    \put {\D\H} at 7 2
    \setdots<2pt>
    \plot 2.4 1  3.6 1 /
    \plot 4.4 1  5.6 1 /
    \plot 1.4 2  2.6 2 /
    \plot 3.4 2  4.6 2 /
    \plot 5.4 2  6.6 2 /
    \plot 2.4 3  3 3 /
    \plot 4.4 3  5.6 3 /
    \plot 3.4 4  4 4 /
    \plot 2.3 1.3  2.7 1.7 /
    \plot 4.3 1.3  4.4 1.4 /
    \plot 6.3 1.3  6.7 1.7 /
    \plot 1.3 2.3  1.7 2.7 /
    \plot 3.3 2.3  3.7 2.7 /
    \plot 5.3 2.3  5.7 2.7 /
    \plot 2.3 3.3  2.7 3.7 /
    \plot 4.3 3.3  4.7 3.7 /
    \plot 1.3 1.7  1.7 1.3 /
    \plot 3.3 1.7  3.7 1.3 /
    \plot 5.3 1.7  5.7 1.3 /
    \plot 2.3 2.7  2.7 2.3 /
    \plot 4.3 2.7  4.4 2.6 /
    \plot 6.3 2.7  6.7 2.3 /
    \plot 3.3 3.7  3.5 3.5 /
    \plot 5.3 3.7  5.7 3.3 /
    \setsolid
    \put {\Narc} at 4 2
    \put {\ding{202},\ding{205}} at 4.9 2.6
    \put {\NWarc} at 4 3 
    \put {\ding{203}} at 4.4 4.2 
    \put {\Sarc} at 4 2 
    \put {\ding{204}} at 4.7 1.4 
    \endpicture
$}
\caption{The basic sequence in the infinite 2:3:4 {\it Tonnetz}}
\label{figure-basic}
\end{figure}

\empar{An example}
Here are  bars 3 and 4 from our sample piece which is reprinted in
Appendix~\ref{appendix-ave}. 
The right hand plays the four chords from the basic sequence while
the left hand emphasizes the tritave-based structure
by repeating the upper two notes of each chord one tritave lower.
The voice in the top line consists of notes taken from those chords.

\begin{figure}[ht]
\begin{center}
  \includegraphics[width=\linewidth]{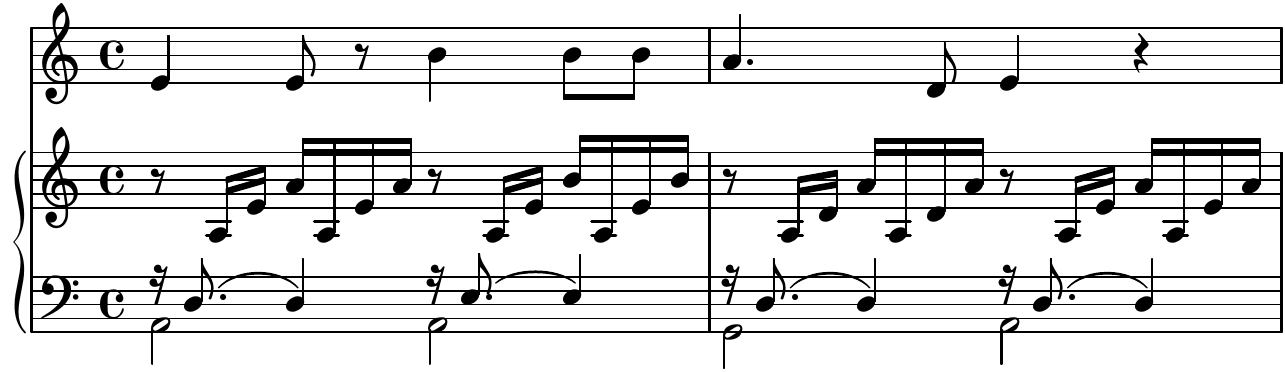}
\end{center}
\label{figure-ave34}
\caption{Bars 3 and 4 from the Ave}
\end{figure}

\subsection{Inversions}
\label{section-inversions}

The term {\it inversion} refers to first or second inversion in which the
bottom note of a chord is substituted by a new top note, or conversely
(here, the term inversion does not refer to mirror inversion).
In octave-based harmony, the two notes differ by an octave, so we speak of
inversions modulo the octave.  In the 2:3:4 system, we use inversions modulo
the tritave.

\empar{Major and minor chords}
We observe that the role of major and minor chords is quite
different in 2:3:4 harmony, when compared to 4:5:6 harmony.

There are three intervals which may occur in a --- possibly inverted --- major or minor triad
in 4:5:6 harmony:  a major third (\ding{195}), a minor third (\ding{194})
or a fourth (\ding{196}).

There are six possibilities to pick two different intervals from the above.
Up to inversion, there are two types:

\smallskip
\centerline{\ding{195}+\ding{194} (major) ---
\ding{194}+\ding{196} (1st inv.\ major) ---
\ding{196}+\ding{195} (2nd inv.\ major)}

\smallskip\noindent
or

\smallskip
\centerline{\ding{194}+\ding{195} (minor) ---
\ding{195}+\ding{196} (1st inv.\ minor) --- \ding{196}+\ding{194}
(2nd inv.\ minor).}

\smallskip\noindent
Tritave-based 2:3:4 harmony is different.  The 2:3:4 chord consists
of a fifth (\ding{198}) and a fourth (\ding{196}), but to complement
the chord to a tritave, another fifth (\ding{198}) is needed.

Hence there are only three possibilities to pick two intervals from the above:

\smallskip
\centerline{\ding{198}+\ding{196} (tonic) ---
\ding{196}+\ding{198} (1st inv.) ---
\ding{198}+\ding{198} (2nd inv.).}

\smallskip\noindent
Note that the first inversion appears as a minor chord,
while the second inversion is an augmented chord.

\empar{A remark regarding perception}
Octave-based 4:5:6 harmony appears to be
richer as there are two modes
(major and minor) and three inversions;
by comparison in tritave-based 2:3:4 harmony, there
is only one mode with two inversions,
or --- looking at it differently ---
three modes (major, minor, augmented) but no inversions.
It would be interesting to find out if a listener perceives the
three chords \ding{198}+\ding{196}, \ding{196}+\ding{198}, and \ding{198}+\ding{198}
as being inversions of the same mode, or rather as representing three different
modes.

Acoustically, that is, when considering the overtone sequences of notes,
some evidence seems to support the first interpretation
(one mode with two inversions).
Consider two numbers associated to a chord:
one is the distance $d_B$ from a base note
for which all notes in the chord are overtones.
For the 2:3:4 chord \A-\E-\A\p, the base note is an octave
below the \A, which is the \A\c\ or \E\V, so $d_B=2$ as in ``2:3:4".

The second measure is the distance $d_O$ to the first common overtone for all the notes
in the chord.  For this, we rewrite the chord using reciprocals with
simplified denominators:  the 2:3:4 chord is the $\frac16:\frac14:\frac13$
chord.  From this notation we see that the first common overtone of the
\A-\E-\A\p\ chord is the \A\p\H\ or \E\p\p, which is one tritave above the \A\p,
hence $d_O=3$ as in ``$\frac13$''.

In the table in Appendix~\ref{subsection-appendix-purity} we list for each chord both numbers.
It turns out that for major 4:5:6 chords, $d_O$ is large and $d_B$ is small;
for minor 4:5:6 chords, $d_O$ is small and $d_B$ is large.
But for each of the three 2:3:4 chords
(\ding{198}+\ding{196}, \ding{196}+\ding{198}, and \ding{198}+\ding{198}),
both numbers are small.

\empar{Inversions in the \boldit{Tonnetz}}
In 4:5:6 harmony, inversions are difficult to picture in the {\it Tonnetz:}  in the finite {\it Tonnetz,}
octave-equivalent notes are identified, hence inversions modulo the octave are invisible.
In the infinite 4:5:6 {\it Tonnetz,} it is not clear which shift
to use for the octave (for example, 3 major thirds or 4 minor thirds?),
but in each case, the inverted chord is no longer contiguous.

\begin{figure}[ht]
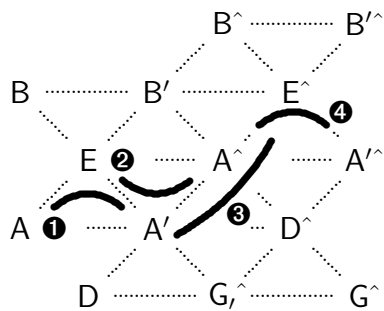

    \centerline{$
    \beginpicture
    \setcoordinatesystem units <5ex,5ex>
    \multiput{} at 2 5  7 1 /
    \put {\D} at 3 1 
    \put {\A} at   2 2
    \put {\G\c\H} at 5 1 
    \put {\A\p} at 4 2 
    \put {\E} at 3 3
    \put {\B} at 2 4 
    \put {\D\H} at 6 2 
    \put {\A\H} at   5 3
    \put {\B\p} at   4 4 
    \put {\G\H} at  7 1 
    \put {\A\p\H} at 7 3 
    \put {\E\H} at 6 4
    \put {\B\H} at 5 5
    \put {\B\p\H} at 7 5     
    \put {\Narc} at 3 2
    \put {\ding{202}} at 2.5 2
    \put {\Sarc} at 4 3
    \put {\ding{203}} at 3.5 3
    \put {\SEarc} at 4.8 2.8
    \put {\ding{204}} at 5.2 2.2
    \put {\Narc} at 6 3.2
    \put {\ding{205}} at 6.7 3.7
    \setdots<2pt>
    \plot 3.4 1  4.6 1 /
    \plot 5.4 1  6.6 1 /
    \plot 3 2  3.6 2 /
    \plot 5.4 2  5.6 2 /
    \plot 4 3  4.6 3 /
    \plot 6 3  6.6 3 /
    \plot 2.4 4  3.6 4 /
    \plot 4.4 4  5.6 4 /
    \plot 5.4 5  6.6 5 /
    \plot 3.3 1.3  3.7 1.7 /
    \plot 5.3 1.3  5.7 1.7 /
    \plot 2.3 2.3  2.7 2.7 /
    \plot 4.3 2.3  4.7 2.7 /
    \plot 6.3 2.3  6.7 2.7 /
    \plot 3.3 3.3  3.7 3.7 /
    \plot 5.3 3.3  5.4 3.4 /
    \plot 4.3 4.3  4.7 4.7 /
    \plot 6.3 4.3  6.7 4.7 /
    \plot 4.3 1.7  4.7 1.3 /
    \plot 6.3 1.7  6.7 1.3 /
    \plot 3.3 2.7  3.7 2.3 /
    \plot 5.5 2.5  5.7 2.3 /
    \plot 2.3 3.7  2.7 3.3 /
    \plot 4.3 3.7  4.7 3.3 /
    \plot 6.6 3.4  6.7 3.3 /
    \plot 5.3 4.7  5.7 4.3 /
    \endpicture$}
\caption{A 2:3:4 chord with three successive inversions}
\label{figure-inversions}
\end{figure}

By comparison, inversions are faithfully represented
in the infinite 2:3:4 {\it Tonnetz.}
Starting from the major \A-\E-\A\p\ chord (\ding{202})
in Figure~\ref{figure-inversions},
the first inversion is the minor  \E-\A\p-\A\H\ chord (\ding{203}).
Inverting again leads to the
augmented chord \A\p-\A\H-\E\H\ (\ding{204}); a third inversion brings us to the
\A\H-\E\H-\A\p\H\ chord (\ding{205}), 
a tritave higher than the chord we started with.

We briefly discuss the role of inversions in our sample piece.

Note that throughout the score of the Ave, the pitch increases by one tritave.
We compare the four chords in bars 3-4 in Figure~\ref{figure-ave34} played by the right hand
with those in bars 9-10, in bars 15-16, and in bars 21-22.

\begin{figure}[ht]
\begin{center}
  \includegraphics[width=\linewidth]{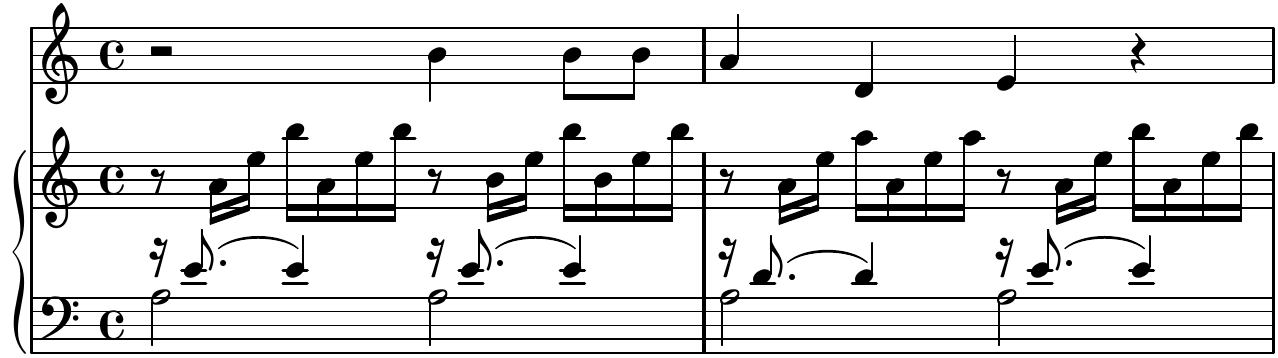}
\end{center}
\label{figure-ave1516}
\caption{Bars 15 and 16 from the Ave}
\end{figure}

The chords in bars 3-4 are the basic sequence, shown in
Figure~\ref{figure-basic}.  Each of those chords is played in first
inversion in the corresponding chord in bars 9-10.
Another first inversion yields the chord sequence in bars 15-16
(see Figure~\ref{figure-ave1516} for the score);
finally, the basic sequence is played one tritave higher in bars 21-22 (in variation).
The first chords in each pair of bars are pictured
in Figure~\ref{figure-inversions}.

Thus the sequence of first inversions in bars 9-10, and the sequence of
second inversions in bars 15-16 divide the Ave into three parts.
They are intended to correspond to the three parts of the lyrics
(greeting; adoration; prayer request).

\subsection{Diminished chords}
\label{subsection-diminished}

In our sample piece, we use diminished chords to move
from a given chord to its inversions in a variety of ways.
As mentioned in the previous section,
the inversions of the chords in the basic sequence
divide the Ave into three parts.  In this subsection, we discuss the third part.

The first and the last chord in the basic sequence is the \A-\E-\A\p,
so the third part of the Ave lies between the second inversion of this chord,
which is the augmented chord \A\p-\A\H-\E\H\ (\ding{202}) in bars 15a
and 16b, see Figure~\ref{figure-ave1516},
and the chord one tritave higher, \A\H-\E\H-\A\p\H\ (\ding{211})
in bars 21a and 22b.

    \begin{figure}[ht]
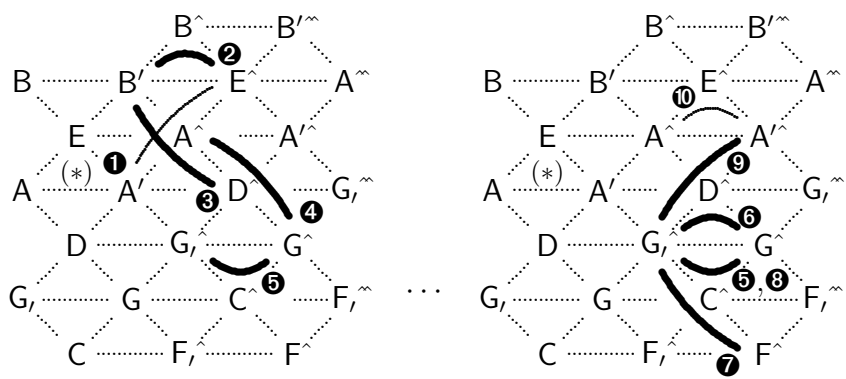

    \centerline{$
    \beginpicture
    \setcoordinatesystem units <4ex,4ex>
    \multiput{} at 2 5  8 -1 /
    %
    \put {\C} at 3 -1
    \put {\G\c} at 2 0 
    \put {\D} at 3 1 
    \put {\A} at   2 2
    \put {\G} at  4 0 
    \put {\F\c\H} at  5 -1 
    \put {\F\H} at 7 -1
    \put {\C\H} at 6 0
    \put {\G\c\H} at 5 1 
    \put {\A\p} at 4 2 
    \put {\E} at 3 3
    \put {\B} at 2 4 
    \put {\D\H} at 6 2 
    \put {\A\H} at   5 3
    \put {\B\p} at   4 4 
    \put {\G\H} at  7 1 
    \put {\F\c\H\H} at  8 0
    \put {\G\c\H\H} at 8 2
    \put {\A\p\H} at 7 3 
    \put {\E\H} at 6 4
    \put {\B\H} at 5 5
    \put {\A\H\H} at   8 4 
    \put {\B\p\H\H} at   7 5 
%
    \put {\footnotesize $(*)$} at 3 2.3
    {\def\plotsym{.}%
    \put {\NWarc} at 5 3 }%
    \put {\ding{202}} at 3.7 2.5
    \put {\Narc} at 5 4
    \put {\ding{203}} at 5.8 4.5
    \put {\SWarc} at 5 3
    \put {\ding{204}} at 5.4  1.8
    \put {\NEarc} at 6 2
    \put {\ding{205}} at 7.3  1.6
    \put {\Sarc} at 6 1
    \put {\ding{206}} at 6.6 .3
    \setdots<2pt>
    \plot 3.4 -1  4.6 -1 /
    \plot 5.4 -1  6.6 -1 /
    \plot 2.4 0  3.6 0 /
    \plot 4.4 0  5.6 0 /
    \plot 7 0  7.6 0 /
    \plot 3.4 1  4.6 1 /
    \plot 5.4 1  6.6 1 /
    \plot 2.4 2  3.6 2 /
    \plot 4.4 2  5 2 /
    \plot 7 2  7.6 2 /
    \plot 3.4 3  4 3 /
    \plot 6 3  6.6 3 /
    \plot 2.4 4  3.6 4 /
    \plot 4.4 4  5.6 4 /
    \plot 6.4 4  7.6 4 /
    \plot 5.4 5  6.6 5 /
    \plot 3.3 -.7  3.7 -.3 /
    \plot 5.3 -.7  5.7 -.3 /
    \plot 7.3 -.7  7.7 -.3 /
    \plot 2.3 .3  2.7 .7 /
    \plot 4.3 .3  4.7 .7 /
    \plot 6.6 .6  6.7 .7 /
    \plot 3.3 1.3  3.7 1.7 /
    \plot 5.3 1.3  5.7 1.7 /
    \plot 2.3 2.3  2.7 2.7 /
    \plot 4.3 2.3  4.5 2.5 /
    \plot 6.5 2.5  6.7 2.7 /
    \plot 3.3 3.3  3.7 3.7 /
    \plot 5.3 3.3  5.7 3.7 /
    \plot 7.3 3.3  7.7 3.7 /
    \plot 4.3 4.3  4.7 4.7 /
    \plot 6.3 4.3  6.7 4.7 /
    \plot 2.3 -.3  2.7 -.7 /
    \plot 4.3 -.3  4.7 -.7 /
    \plot 6.3 -.3  6.7 -.7 /
    \plot 3.3 .7  3.7 .3 /
    \plot 5.3 .7  5.7 .3 /
    \plot 7.3 .7  7.7 .3 /
    \plot 2.3 1.7  2.7 1.3 /
    \plot 4.3 1.7  4.7 1.3 /
    \plot 6.3 1.7  6.7 1.3 /
    \plot 3.3 2.7  3.4 2.6 /
    \plot 5.3 2.7  5.7 2.3 /
    \plot 7.3 2.7  7.7 2.3 /
    \plot 2.3 3.7  2.7 3.3 /
    \plot 4.3 3.7  4.5 3.5 /
    \plot 6.3 3.7  6.7 3.3 /
    \plot 5.3 4.7  5.5 4.5 /
    \plot 7.3 4.7  7.7 4.3 /
    \endpicture
    \quad\cdots\quad
    \beginpicture
    \setcoordinatesystem units <4ex,4ex>
    \multiput{} at 2 5  8 -1 /
    %
    \put {\C} at 3 -1
    \put {\G\c} at 2 0 
    \put {\D} at 3 1 
    \put {\A} at   2 2
    \put {\G} at  4 0 
    \put {\F\c\H} at  5 -1 
    \put {\F\H} at 7 -1
    \put {\C\H} at 6 0
    \put {\G\c\H} at 5 1 
    \put {\A\p} at 4 2 
    \put {\E} at 3 3
    \put {\B} at 2 4 
    \put {\D\H} at 6 2 
    \put {\A\H} at   5 3
    \put {\B\p} at   4 4 
    \put {\G\H} at  7 1 
    \put {\F\c\H\H} at  8 0
    \put {\G\c\H\H} at 8 2
    \put {\A\p\H} at 7 3 
    \put {\E\H} at 6 4
    \put {\B\H} at 5 5
    \put {\A\H\H} at   8 4 
    \put {\B\p\H\H} at   7 5 
    %
    \put {\footnotesize $(*)$} at 3 2.3
    \put {\Narc} at 6 1
    \put {\ding{207}} at 6.7 1.5
    \put {\SWarc} at 6 0
    \put {\ding{208}} at 6.3 -1.2
    \put {\Sarc} at 6 1
    \put {\ding{206},\ding{209}} at 6.9 .3
    \put {\NWarc} at 6 2
    \put {\ding{210}} at 6.5 2.5
    {\def\plotsym{.}%
    \put {\Narc} at 6 3 }%
    \put {\ding{211}} at 5.5 3.7
    \setdots<2pt>
    \plot 3.4 -1  4.6 -1 /
    \plot 5.4 -1  6 -1 /
    \plot 2.4 0  3.6 0 /
    \plot 4.4 0  5 0 /
    \plot 6.4 0  7.6 0 /
    \plot 3.4 1  4.6 1 /
    \plot 5.4 1  6.6 1 /
    \plot 2.4 2  3.6 2 /
    \plot 4.4 2  5 2 /
    \plot 6.4 2  7.6 2 /
    \plot 3.4 3  4.6 3 /
    \plot 5.4 3  6.6 3 /
    \plot 2.4 4  3.6 4 /
    \plot 4.4 4  5.6 4 /
    \plot 6.4 4  7.6 4 /
    \plot 5.4 5  6.6 5 /
    \plot 3.3 -.7  3.7 -.3 /
    \plot 5.3 -.7  5.5 -.5 /
    \plot 7.3 -.7  7.7 -.3 /
    \plot 2.3 .3  2.7 .7 /
    \plot 4.3 .3  4.7 .7 /
    \plot 6.6 .6  6.7 .7 /
    \plot 3.3 1.3  3.7 1.7 /
    \plot 5.3 1.3  5.7 1.7 /
    \plot 7.3 1.3  7.7 1.7 /
    \plot 2.3 2.3  2.7 2.7 /
    \plot 4.3 2.3  4.7 2.7 /
    \plot 6.3 2.3  6.4 2.4 /
    \plot 3.3 3.3  3.7 3.7 /
    \plot 5.3 3.3  5.4 3.4 /
    \plot 7.3 3.3  7.7 3.7 /
    \plot 4.3 4.3  4.7 4.7 /
    \plot 6.3 4.3  6.7 4.7 /
    \plot 2.3 -.3  2.7 -.7 /
    \plot 4.3 -.3  4.7 -.7 /
    \plot 6.3 -.3  6.7 -.7 /
    \plot 3.3 .7  3.7 .3 /
    \plot 5.3 .7  5.7 .3 /
    \plot 7.3 .7  7.7 .3 /
    \plot 2.3 1.7  2.7 1.3 /
    \plot 4.3 1.7  4.7 1.3 /
    \plot 6.3 1.7  6.4 1.6 /
    \plot 3.3 2.7  3.7 2.3 /
    \plot 5.3 2.7  5.5 2.5 /
    \plot 7.3 2.7  7.7 2.3 /
    \plot 2.3 3.7  2.7 3.3 /
    \plot 4.3 3.7  4.7 3.3 /
    \plot 6.3 3.7  6.7 3.3 /
    \plot 5.3 4.7  5.7 4.3 /
    \plot 7.3 4.7  7.7 4.3 /
    \endpicture$}
    \caption{{\it Tonnetz} for Ave, bars 16b -- 21a}
    \label{figure-ave1621}
  \end{figure}

\begin{figure}[ht]
\begin{center}
  \includegraphics[width=\linewidth]{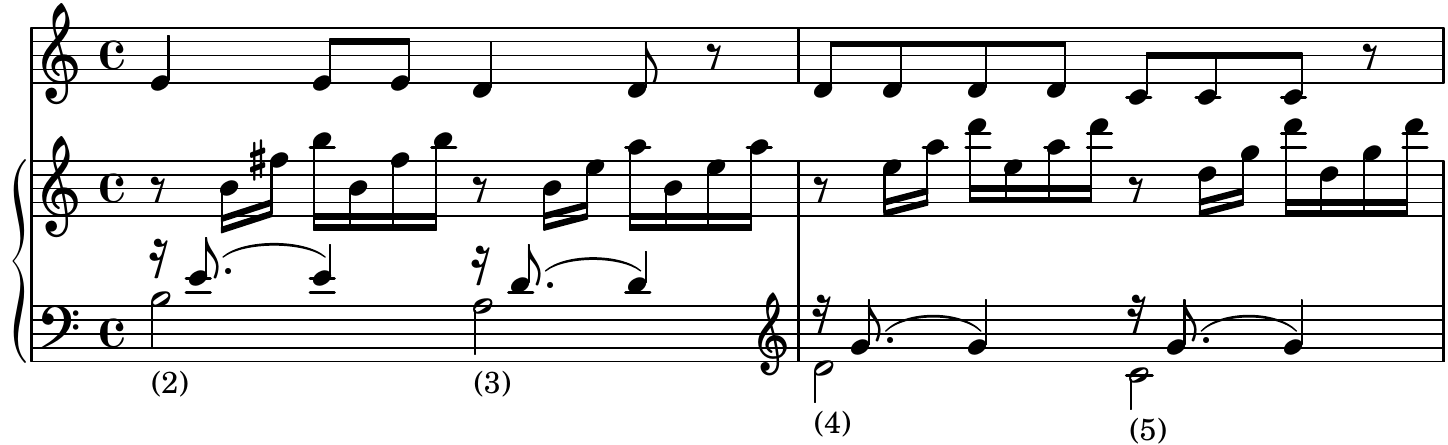}

  \includegraphics[width=\linewidth]{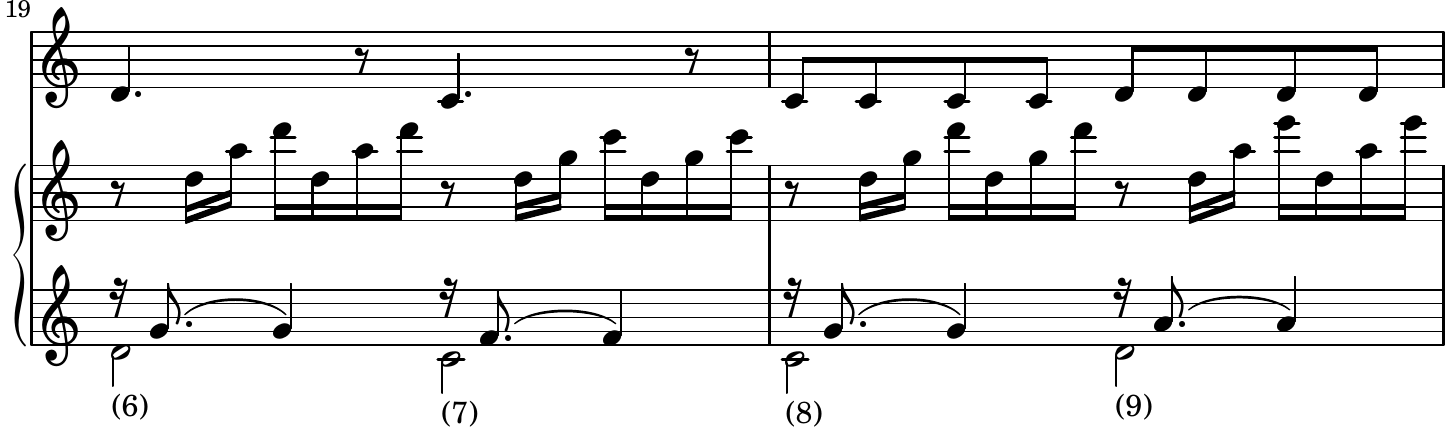}
\end{center}
\label{figure-ave1720}
\caption{Bars 17 to 20 from the Ave}
\end{figure}

Figure~\ref{figure-ave1621} shows the {\it Tonnetz} for the third
part of the Ave; in Figure~\ref{figure-ave1720} we reprint the corresponding
bars 17 - 20.

In bars 17 - 18, we start from the 
major chord (\ding{203}), then move
to the diminished chord \B\p-\A\H-\D\H\ (\ding{204}),
push  the diminished chord up to \A\H-\D\H-\G\H\ (\ding{205})
and resolve to minor \G\c\H-\C\H-\G\H\ (\ding{206}).
The S-shaped figure \ding{203}-\ding{206} contains the notes \E\H, \D\H, \C\H;
hence the notes played by the left hand support the melody \E-\D-\C\ 
one tritave lower.

The third part of the Ave concludes with bars 19 - 20.

The minor chord (\ding{206}) is followed by its subdominant (\ding{207}),
then the harmony descends
in the {\it Tonnetz}  to the diminished triad (\ding{208}),
returns to the minor (\ding{209}), moves up to its second subdominant
(\ding{210}), and continues the upwards movement 
to the second subdominant for the previous chord,
which is the final major chord \A\H-\E\H-\A\p\H\ (\ding{211}).

Throughout this part, the melody line has W-shape (\E-\D-\C-\D-\C-\D-\E)
which is reflected in the up-and-down movement of the harmony in the {\it Tonnetz}
in Figure~\ref{figure-ave1621}.

\subsection{Purity and sparsity}
\label{subsection-purity}

When asking listeners about their impression of the
{\it Ave Maria in dix-neuf par duodecime,} the piece has been
characterized as ``reminding of Gregorian chants'', as
``sounding Celtic'',
and as having a sound that is ``pure and sparse''.
Hence this section on purity and sparsity.

\empar{Purity}
For a chord, two measures for purity come to mind; the numbers
$d_O$ and $d_B$ from Section~\ref{section-inversions} measure
the distance between the highest note in the chord and the first common overtone,
and the distance between the lowest note and a base note for which all
notes in the chord are overtones, respectively.

In Appendix~\ref{subsection-appendix-purity} we compute both measures
for the 4:5:6 system and for 2:3:4 harmony.
Major chords in both systems
stand out for their small distance from the base note, while
minor chords 
are very close to the first common overtone.  Also in both systems,
augmented chords do better in both measures than diminished chords.

Overall, all chords 
are much purer with respect to both measures in 2:3:4 harmony than in the
4:5:6 system.
On the other hand, as we have discussed in Section~\ref{section-inversions},
there may be less variety in 2:3:4 harmony as the distinction between
major and minor chords is less pronounced.

\empar{Sparsity}
It is possible that the impression of sparsity arises since the 
intervals 5:4 and 6:5 in 4:5:6 harmony are missing in
the 2:3:4 system (because any frequency ratio between two notes
has the form $2^u\cdot 3^v$, so no prime factor 5 occurs).
With the major and minor third missing, chords are spaced further apart
and hence may sound ``thinner''.

The other reason for sparsity is that for a piece in 2:3:4 harmony
it is necessary to emphasize the tritave to create the proper frame.
Throughout the {\it Ave} for example, the left hand
reproduces the top two notes of the right hand, but one tritave lower.
Clearly, by playing notes in parallel, the density of overtones is reduced,
which the listener may perceive as pure yet sparse.

\empar{Further comments regarding perception}
\begin{itemize}
\item We have seen in Section~\ref{section-neo-Riemannian} that harmonic development
along fundamental moves in the {\it Tonnetz} limits the voice to fewer options than in the 4:5:6 system,
perhaps adding to the piece being perceived as pure yet sparce.

\item The author would find it interesting to study perception of chord sequences
in 2:3:4 harmony in a project similar to the research in \cite{walker} for the
Bohlen-Pierce scale.

\item
Since the 2:3:4 chord 
consists of a fourth, a fifth, and an octave, we cannot omit
a quote from the text book \cite[Section 3.8.2]{loy}
with which we conclude this section.

\begin{quote}
The interval of the third in the Pythagorean scale
was considered a dissonance in the Middle Ages, and as a result
compositions would typically omit the third in the final chord
of a composition so as to end only with perfect intervals ---
fourths, fifths, and octaves --- an effect that sounds hollow
to modern ears.
\end{quote}
\end{itemize}

\section*{Acknowledgements}
Thanks to Elaine Walker for her guidance and advice,
in particular regarding the relevance and the role
of the frame given by the octave or the tritave. 
The author is grateful for many helpful discussions with Feruza Dadabaeva, his piano teacher.
He would like to thank Associate Editor Emmanuel Amiot and Co-Editor-in-Chief 
Jason Yust and the referees from the Journal of Mathematics and Music,
for comments which have led to substantial 
improvements of the manuscript.



\FloatBarrier

\section{Appendices}

\subsection{Comparison of Pythagorean vs.\ equally tempered scales}
\label{appendix-a}

For the Pythagorean scales Pyth-2 and Pyth-3, we list in Tables~\ref{table-comparison} and
\ref{table-19edt} for each note the scale degree,
the harmonic degree, the frequencies associated in just and in equal intonation,
and their quotient, measured in cents.

Both scales satisfy the symmetry condition in \cite{cc} and hence can be considered well-formed.
In \cite[Table~2]{cc},
Pyth-2 is the chromatic scale with group $\mathbb Z_N=\mathbb Z_{12}$ in which the fifth has scale degree
$b=7$.  The scale Pyth-3 is not listed there explicitely (since the scale generator is an octave, not a fifth);
it is based on the group $\mathbb Z_N=\mathbb Z_{19}$ in which the octave, which is the scale generator,
has degree $b=12$. Hence in Table~\ref{table-19edt}, the scale degree is obtained from the harmonic degree
by multiplication by $b=12$, for example harmonic degrees $0, 1, 2,\ldots$ correspond
to scale degrees $0, 12\equiv -7({\rm mod}\; 19), 24\equiv 5({\rm mod}\; 19), \ldots$
Conversely, since $b^{-1}\equiv 8({\rm mod}\; 19)$, scale degrees $0, 1, 2,\ldots$ correspond to harmonic degrees
$0, 8, 16\equiv-3({\rm mod}\; 19),\ldots$

\begin{table}
\caption{Two octave-based scales: Pyth-2 vs.\ 12-EDO}
{\begin{tabular}{|c|c|r@{\,=\,}l|c||r@{\,$\approx$\,}l|c|} \hline
    \multicolumn{8}{|c|}{\bfseries Pythagorean vs.\ 12 tone equal tuning}\\ \hline
    \multicolumn2{|c|}{\it scale} & \multicolumn2{|c|}{\rm Pyth-2} &  {\it harm.} 
                  & \multicolumn2{|c|}{\rm 12-EDO} & $\pi(n)/2^{\frac{n}{12}}$ \\ [-1ex]
    \multicolumn2{|c|}{\it degree} & \multicolumn2{|c|}{\it tuning} & {\it degree} 
    & \multicolumn2{|c|}{\it tuning} & {\it in} \\ [-1ex]
    \multicolumn2{|c|}{$n$} & \multicolumn2{|c|}{$\pi(n)$} & $\mu_3(\pi(n))$
                 & \multicolumn2{|c|}{$2^{\frac n{12}}$} & {\it cents} \\ \hline \hline
    (-6) & \A\f   & $\frac{512}{729}$ & $2^{9}\cdot 3^{-6}$ & $-6$
                  & $2^{-\frac{6}{12}}$ & 0.7071 & $-11.73$ \\ \hline
    -5 & \A       & $\frac{3}{4}$ & $2^{-2}\cdot 3^{1}$ & 1              
                  & $2^{-\frac{5}{12}}$ & 0.7492 & $+1.96$ \\ \hline
    -4 & \B\f     & $\frac{64}{81}$ & $2^{6}\cdot 3^{-4}$ & $-4$
                  & $2^{-\frac{4}{12}}$ & 0.7937 & $-7.82$ \\ \hline
    -3 & \B       & $\frac{27}{32}$ & $2^{-5}\cdot 3^{3}$ & 3              
                  & $2^{-\frac{3}{12}}$ & 0.8409 & $+5.87$ \\ \hline
    -2 & \C       & $\frac{8}{9}$ & $2^{3}\cdot 3^{-2}$ & $-2$
                  & $2^{-\frac{2}{12}}$ & 0.8909 & $-3.91$ \\ \hline
    -1 & \C\s     & $\frac{243}{256}$ & $2^{-8}\cdot 3^{5}$ & 5              
                  & $2^{-\frac{1}{12}}$ & 0.9439 & $+9.78$ \\ \hline
    0  & \D       & $1$ & $2^{0}\cdot 3^{0}$ & $0$
                  & $2^0$ & 1.0000 & $\pm0$ \\ \hline
    1  & \E\f     & $\frac{256}{243}$ & $2^{8}\cdot 3^{-5}$ & $-5$              
                  & $2^{\frac{1}{12}}$ & 1.0595 & $-9.78$ \\ \hline
    2  & \E       & $\frac{9}{8}$ & $2^{-3}\cdot 3^{2}$ & $2$
                  & $2^{\frac{2}{12}}$ & 1.1225 & $+3.91$ \\ \hline
    3  & \F       & $\frac{32}{27}$ & $2^{5}\cdot 3^{-3}$ & $-3$              
                  & $2^{\frac{3}{12}}$ & 1.1892 & $-5.87$ \\ \hline
    4  & \F\s     & $\frac{81}{64}$ & $2^{-6}\cdot 3^{4}$ & 4              
                  & $2^{\frac{4}{12}}$ & 1.2599 & $+7.82$ \\ \hline
    5 & \G &      $\frac{4}{3}$ & $2^{2}\cdot 3^{-1}$ & $-1$              
                  & $2^{\frac{5}{12}}$ & 1.3348 & $-1.96$ \\ \hline
    6  & \G\s     & $\frac{729}{512}$ & $2^{-9}\cdot 3^{6}$ & 6              
                  & $2^{\frac{6}{12}}$ & 1.4142 & $+11.73$ \\ \hline
  \end{tabular}}
  \label{table-comparison}
\end{table}

\begin{table}
\caption{Two tritave-based scales:  Pyth-3 and 19-EDT}
{\begin{tabular}{|r|c|r@{\,=\,}l|c||r@{\,=\,}l|c|} \hline
    \multicolumn{8}{|c|}{\bfseries Pythagorean vs.\ 19 tones equal in the tritave}\\ \hline
    \multicolumn2{|c|}{\it scale} & \multicolumn2{|c|}{\rm Pyth-3} &  {\it harm.} 
    & \multicolumn2{|c|}{\rm 19-EDT} & $\pi_3(n)/3^{\frac n{19}}$ \\ [-1ex]
    \multicolumn2{|c|}{\it degree} & \multicolumn2{|c|}{\it tuning} & {\it degree} 
    & \multicolumn2{|c|}{\it tuning} & {\it in} \\ [-1ex]
    \multicolumn2{|c|}{$n$} & \multicolumn2{|c|}{$\pi_3(n)$} & $\mu_2(\pi_3(n))$
                 & \multicolumn2{|c|}{$3^{\frac n{19}}$} &  {\it cents} \\ \hline \hline
    (-10) & \E\c  & $\frac{9}{16}$ & $2^{-4}\cdot 3^{2}$ & $-4$
                  & $3^{-\frac{10}{19}}$ & 0.5609 & $+4.94$ \\ \hline
    -9 & \F\c     & $\frac{16}{27}$ & $2^{4}\cdot 3^{-3}$ & $4$              
                  & $3^{-\frac{9}{19}}$ & 0.5943 & $-4.94$ \\ \hline
    -8 & \F\s\c   & $\frac{81}{128}$ & $2^{-7}\cdot 3^{4}$ & $-7$
                  & $3^{-\frac{8}{19}}$ & 0.6297 & $+8.64$ \\ \hline
    -7 & \G\c     & $\frac{2}{3}$ & $2^{1}\cdot 3^{-1}$ & $1$            
                  & $3^{-\frac{7}{19}}$ & 0.6671 & $-1.23$ \\ \hline
    -6 & \A\f     & $\frac{512}{729}$ & $2^{9}\cdot 3^{-6}$ & $9$
                  & $3^{-\frac{6}{19}}$ & 0.7069 & $-11.11$ \\ \hline
    -5 & \A       & $\frac{3}{4}$ & $2^{-2}\cdot 3^{1}$ & $-2$              
                  & $3^{-\frac{5}{19}}$ & 0.7489 & $+2.47$ \\ \hline
    -4 & \B\f     & $\frac{64}{81}$ & $2^{6}\cdot 3^{-4}$ & $6$
                  & $3^{-\frac{4}{19}}$ & 0.7935 & $-7.41$ \\ \hline
    -3 & \B       & $\frac{27}{32}$ & $2^{-5}\cdot 3^{3}$ & $-5$  
                  & $3^{-\frac{3}{19}}$ & 0.8408 & $+6.17$ \\ \hline
    -2 & \C       & $\frac{8}{9}$ & $2^{3}\cdot 3^{-2}$ & $3$
                  & $3^{-\frac{2}{19}}$ & 0.8908 & $-3.70$ \\ \hline
    -1 & \C\s     & $\frac{243}{256}$ & $2^{-8}\cdot 3^{5}$ & $-8$              
                  & $3^{-\frac{1}{19}}$ & 0.9438 & $+9.88$ \\ \hline
    0  & \D       & $1$ & $2^{0}\cdot 3^{0}$ & $0$
                  & $3^0$ & 1.0000 & $\pm0$ \\ \hline
    1  & \E\f     & $\frac{256}{243}$ & $2^{8}\cdot 3^{-5}$ & $8$              
                  & $3^{\frac{1}{19}}$ & 1.0595 & $-9.88$ \\ \hline
    2  & \E       & $\frac{9}{8}$ & $2^{-3}\cdot 3^{2}$ & $-3$
                  & $3^{\frac{2}{19}}$ & 1.1226 & $+3.70$ \\ \hline
    3  & \F       & $\frac{32}{27}$ & $2^{5}\cdot 3^{-3}$ & $5$              
                  & $3^{\frac{3}{19}}$ & 1.1894 & $-6.17$ \\ \hline
    4  & \F\s     & $\frac{81}{64}$ & $2^{-6}\cdot 3^{4}$ & $-6$            
                  & $3^{\frac{4}{19}}$ & 1.2602 & $+7.41$ \\ \hline
    5 & \G        & $\frac{4}{3}$ & $2^{2}\cdot 3^{-1}$ & $2$              
                  & $3^{\frac{5}{19}}$ & 1.3352 & $-2.47$ \\ \hline
    6  & \G\s     & $\frac{729}{512}$ & $2^{-9}\cdot 3^{6}$ & $-9$              
                  & $3^{\frac{6}{19}}$ & 1.4147 & $+11.11$ \\ \hline
    7  & \A\p     & $\frac{3}{2}$ & $2^{-1}\cdot 3^{1}$ & $-1$              
                  & $3^{\frac{7}{19}}$ & 1.4989 & $+1.24$ \\ \hline
    8  & \B\f\p   & $\frac{128}{81}$ & $2^{7}\cdot 3^{-4}$ & 7              
                  & $3^{\frac{8}{19}}$ & 1.5882 & $-8.64$ \\ \hline
    9 & \B\p      & $\frac{27}{16}$ & $2^{-4}\cdot 3^{3}$ & $-4$              
                  & $3^{\frac{9}{19}}$ & 1.6827 & $+4.94$ \\ \hline
    (10)  & \C\p  & $\frac{16}{9}$ & $2^{4}\cdot 3^{-2}$ & 4              
                  & $3^{\frac{10}{19}}$ & 1.7829 & $-4.94$ \\ \hline
  \end{tabular}}
  \label{table-19edt}
\end{table}


\subsection{On the number of notes per tritave}
\label{appendix-number-of-notes}
What is the best number of notes in a scale?  Suppose our system will have $q$ notes per tritave,
and the octave will be at the $p$-th note, $0<p<q$.  Then we have the
approximation
$$2^q\approx 3^p$$
or, equivalently,
$$q \log(2) \approx p \log(3),\quad \text{or}\quad
\frac{\log(2)}{\log(3)}\approx \frac pq.$$
We need to approximate $\log(2)/\log(3)$ by a rational number,
this can be done using continued fractions, see 
\cite[Section~6.2]{benson}, \cite{krantz} and, for an application to
the Bohlen-Pierce scale, \cite[Section~2.6]{advocat}.

$$\frac{\log(2)}{\log(3)}\;=\;\cfrac1{1+\cfrac1{1+\cfrac1{1+\cfrac1{2+
        \cfrac1{2+\cfrac1{3+\cfrac1{1+\cfrac1{5+\cdots}}}}}}}}
	\approx0.630930$$

Using the first five terms (omitting the summand after the second number 2)
yields the approximation
$$\frac{\log(2)}{\log(3)}\approx\frac{12}{19}.$$
It is not a surprise that
this is a good approximation since the number $3^{12}/2^{19}$,
being the Pythagorean comma $\kappa$, is very close to one.
It gives rise to a scale of just intonation of 19 notes per tritave with the
octave at scale degree 12.  This is the scale Pyth-3 studied in Section~1.

Incidentally, the approximation at degree 7 (given by omitting the fraction
containing the 5) is an excellent one:
$\log(2)/\log(3)\approx 53/84$.  It yields a scale of 84 notes
of which the octave is at scale degree 53.  This scale corresponds to the
octave-based scale of 53 notes generated by the fifth at scale degree 31
\cite[3.14.1]{loy}.

\subsection{Key labels for the 88-key piano keyboard}
\label{appendix-labels}

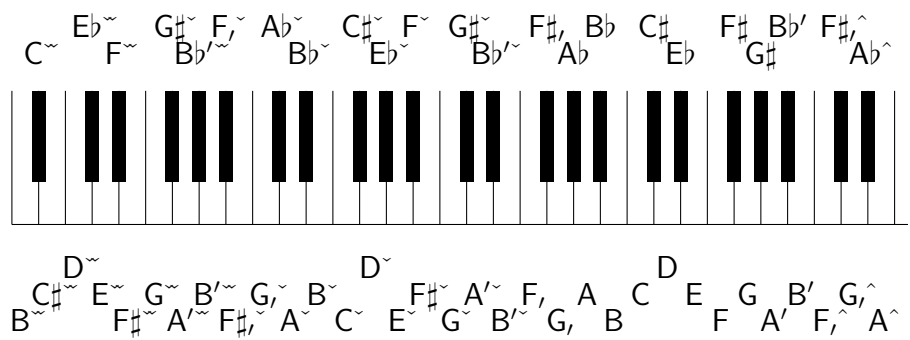
\begin{figure}[ht]
  \begin{center}
    \setlength\unitlength{.14in}
    \begin{picture}(31,12)(0,-4)
      \put(-2,0){\line(1,0){33.5}}
      \multiput(-2,0)(1,0){34}{\line(0,1){5}}
      \linethickness{.07in}
      \multiput(0,0)(7,0)5{\multiput(1,1.6)(1,0)2{\line(0,1){3.4}}}
      \multiput(0,0)(7,0)4{\multiput(4,1.6)(1,0)3{\line(0,1){3.4}}}
      \put(-1,1.6){\line(0,1){3.4}}
      \put(-1.5,-4){\makebox[0ex]{\B\V\V}}
      \put(-.5,-3){\makebox[0ex]{\C\s\V\V}}
      \put(0.5,-2){\makebox[0ex]{\D\V\V}}
      \put(1.5,-3){\makebox[0ex]{\E\V\V}}
      \put(2.5,-4){\makebox[0ex]{\F\s\V\V}}
      \put(3.5,-3){\makebox[0ex]{\G\V\V}}
      \put(4.5,-4){\makebox[0ex]{\A\p\V\V}}
      \put(5.5,-3){\makebox[0ex]{\B\p\V\V}}
      \put(6.5,-4){\makebox[0ex]{\F\s\c\V}}
      \put(7.5,-3){\makebox[0ex]{\G\c\V}}
      \put(8.5,-4){\makebox[0ex]{\A\V}}
      \put(9.5,-3){\makebox[0ex]{\B\V}}
      \put(10.5,-4){\makebox[0ex]{\C\V}}
      \put(11.5,-2){\makebox[0ex]{\D\V}}
      \put(12.5,-4){\makebox[0ex]{\E\V}}
      \put(13.5,-3){\makebox[0ex]{\F\s\V}}
      \put(14.5,-4){\makebox[0ex]{\G\V}}
      \put(15.5,-3){\makebox[0ex]{\A\p\V}}
      \put(16.5,-4){\makebox[0ex]{\B\p\V}}
      \put(17.5,-3){\makebox[0ex]{\F\c}}
      \put(18.5,-4){\makebox[0ex]{\G\c}}
      \put(19.5,-3){\makebox[0ex]{\A}}
      \put(20.5,-4){\makebox[0ex]{\B}}
      \put(21.5,-3){\makebox[0ex]{\C}}
      \put(22.5,-2){\makebox[0ex]{\D}}
      \put(23.5,-3){\makebox[0ex]{\E}}
      \put(24.5,-4){\makebox[0ex]{\F}}
      \put(25.5,-3){\makebox[0ex]{\G}}
      \put(26.5,-4){\makebox[0ex]{\A\p}}
      \put(27.5,-3){\makebox[0ex]{\B\p}}
      \put(28.5,-4){\makebox[0ex]{\F\c\H}}
      \put(29.5,-3){\makebox[0ex]{\G\c\H}}
      \put(30.5,-4){\makebox[0ex]{\A\H}}
      \put(-1,6){\makebox[0ex]{\C\V\V}}
      \put(1,7){\makebox[0ex]{\E\f\V\V}}
      \put(2,6){\makebox[0ex]{\F\V\V}}
      \put(4,7){\makebox[0ex]{\G\s\V}}
      \put(5,6){\makebox[0ex]{\B\f\p\V\V}}
      \put(6,7){\makebox[0ex]{\F\c\V}}
      \put(8,7){\makebox[0ex]{\A\f\V}}
      \put(9,6){\makebox[0ex]{\B\f\V}}
      \put(11,7){\makebox[0ex]{\C\s\V}}
      \put(12,6){\makebox[0ex]{\E\f\V}}
      \put(13,7){\makebox[0ex]{\F\V}}
      \put(15,7){\makebox[0ex]{\G\s\V}}
      \put(16,6){\makebox[0ex]{\B\f\p\V}}
      \put(18,7){\makebox[0ex]{\F\s\c}}
      \put(19,6){\makebox[0ex]{\A\f}}
      \put(20,7){\makebox[0ex]{\B\f}}
      \put(22,7){\makebox[0ex]{\C\s}}
      \put(23,6){\makebox[0ex]{\E\f}}
      \put(25,7){\makebox[0ex]{\F\s}}
      \put(26,6){\makebox[0ex]{\G\s}}
      \put(27,7){\makebox[0ex]{\B\f\p}}
      \put(29,7){\makebox[0ex]{\F\s\c\H}}
      \put(30,6){\makebox[0ex]{\A\f\H}}
    \end{picture}
  \end{center}
    \caption{Key labels for the 88-key keyboard (left part)}
    \label{fig-keys-left}
\end{figure}

  In Figures~\ref{fig-keys-left} and \ref{fig-keys-right}
  we present the key assignment for an 88-key piano keyboard.
  Either equal intonation (12-EDO) or
  just intonation (Pyth-2) can be used to play 19 notes per tritave,
  see Section~\ref{subsection-piano}.

\begin{figure}[ht]
  \begin{center}
    \setlength\unitlength{.14in}
    \begin{picture}(33,12)(2.5,-4)
      \put(2.5,0){\line(1,0){33.5}}
      \multiput(3,0)(1,0){34}{\line(0,1){5}}
      \linethickness{.07in}
      \multiput(7,0)(7,0)4{\multiput(1,1.6)(1,0)2{\line(0,1){3.4}}}
      \multiput(0,0)(7,0)5{\multiput(4,1.6)(1,0)3{\line(0,1){3.4}}}
      \put(3.5,-3){\makebox[0ex]{\F\c}}
      \put(4.5,-4){\makebox[0ex]{\G\c}}
      \put(5.5,-3){\makebox[0ex]{\A}}
      \put(6.5,-4){\makebox[0ex]{\B}}
      \put(7.5,-3){\makebox[0ex]{\C}}
      \put(8.5,-2){\makebox[0ex]{\D}}
      \put(9.5,-3){\makebox[0ex]{\E}}
      \put(10.5,-4){\makebox[0ex]{\F}}
      \put(11.5,-3){\makebox[0ex]{\G}}
      \put(12.5,-4){\makebox[0ex]{\A\p}}
      \put(13.5,-3){\makebox[0ex]{\B\p}}
      \put(14.5,-4){\makebox[0ex]{\F\c\H}}
      \put(15.5,-3){\makebox[0ex]{\G\c\H}}
      \put(16.5,-4){\makebox[0ex]{\A\H}}
      \put(17.5,-3){\makebox[0ex]{\B\f\H}}
      \put(18.5,-4){\makebox[0ex]{\C\H}}
      \put(19.5,-2){\makebox[0ex]{\D\H}}
      \put(20.5,-4){\makebox[0ex]{\E\H}}
      \put(21.5,-3){\makebox[0ex]{\F\H}}
      \put(22.5,-4){\makebox[0ex]{\G\H}}
      \put(23.5,-3){\makebox[0ex]{\A\p\H}}
      \put(24.5,-4){\makebox[0ex]{\B\f\p\H}}
      \put(25.5,-3){\makebox[0ex]{\F\c\H\H}}
      \put(26.5,-4){\makebox[0ex]{\G\c\H\H}}
      \put(27.5,-3){\makebox[0ex]{\A\H\H}}
      \put(28.5,-4){\makebox[0ex]{\B\f\H\H}}
      \put(29.5,-3){\makebox[0ex]{\C\H\H}}
      \put(30.5,-2){\makebox[0ex]{\D\H\H}}
      \put(31.5,-3){\makebox[0ex]{\E\f\H\H}}
      \put(32.5,-4){\makebox[0ex]{\F\H\H}}
      \put(33.5,-3){\makebox[0ex]{\G\H\H}}
      \put(34.5,-4){\makebox[0ex]{\A\p\H\H}}
      \put(35.5,-3){\makebox[0ex]{\B\f\p\H\H}}
      \put(4,7){\makebox[0ex]{\F\s\c}}
      \put(5,6){\makebox[0ex]{\A\f}}
      \put(6,7){\makebox[0ex]{\B\f}}
      \put(8,7){\makebox[0ex]{\C\s}}
      \put(9,6){\makebox[0ex]{\E\f}}
      \put(11,7){\makebox[0ex]{\F\s}}
      \put(12,6){\makebox[0ex]{\G\s}}
      \put(13,7){\makebox[0ex]{\B\f\p}}
      \put(15,7){\makebox[0ex]{\F\s\c\H}}
      \put(16,6){\makebox[0ex]{\A\f\H}}
      \put(18,7){\makebox[0ex]{\B\H}}
      \put(19,6){\makebox[0ex]{\C\s\H}}
      \put(20,7){\makebox[0ex]{\E\f\H}}
      \put(22,7){\makebox[0ex]{\F\s\H}}
      \put(23,6){\makebox[0ex]{\G\s\H}}
      \put(25,7){\makebox[0ex]{\B\p\H}}
      \put(26,6){\makebox[0ex]{\F\s\c\H\H}}
      \put(27,7){\makebox[0ex]{\A\f\H\H}}
      \put(29,7){\makebox[0ex]{\B\f\H\H}}
      \put(30,6){\makebox[0ex]{\C\s\H\H}}
      \put(32,7){\makebox[0ex]{\E\H\H}}
      \put(33,6){\makebox[0ex]{\F\s\H\H}}
      \put(34,7){\makebox[0ex]{\G\s\H\H}}
    \end{picture}
  \end{center}
    \caption{Key labels for the 88-key keyboard (right part)}
    \label{fig-keys-right}
\end{figure}
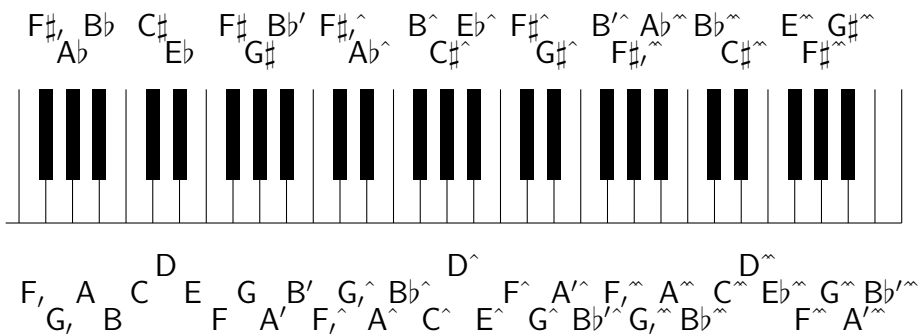

\begin{figure}[!htpb]
\begin{center}
\setlength\unitlength{.14in}
\begin{picture}(33,12)(0,-4)
\put(-.5,0){\line(1,0){34}}
\multiput(0,0)(1,0){34}{\line(0,1){5}}
\linethickness{.07in}
\multiput(0,0)(11,0)3{\multiput(5,1.6)(1,0)2{\line(0,1){3.4}}}
\multiput(0,0)(11,0)3{\multiput(1,1.6)(1,0)3{\line(0,1){3.4}}}
\multiput(0,0)(11,0)3{\multiput(8,1.6)(1,0)3{\line(0,1){3.4}}}
\put(0.5,-3){\makebox[0ex]{\F\c\V}}
\put(1.5,-4){\makebox[0ex]{\G\c\V}}
\put(2.5,-3){\makebox[0ex]{\A\V}}
\put(3.5,-4){\makebox[0ex]{\B\V}}
\put(4.5,-3){\makebox[0ex]{\C\V}}
\put(5.5,-2){\makebox[0ex]{\D\V}}
\put(6.5,-3){\makebox[0ex]{\E\V}}
\put(7.5,-4){\makebox[0ex]{\F\V}}
\put(8.5,-3){\makebox[0ex]{\G\V}}
\put(9.5,-4){\makebox[0ex]{\A\p\V}}
\put(10.3,-3){\makebox[0ex]{\B\p\V}}
\put(11.7,-3){\makebox[0ex]{\F\c}}
\put(12.5,-4){\makebox[0ex]{\G\c}}
\put(13.5,-3){\makebox[0ex]{\A}}
\put(14.5,-4){\makebox[0ex]{\B}}
\put(15.5,-3){\makebox[0ex]{\C}}
\put(16.5,-2){\makebox[0ex]{\D}}
\put(17.5,-3){\makebox[0ex]{\E}}
\put(18.5,-4){\makebox[0ex]{\F}}
\put(19.5,-3){\makebox[0ex]{\G}}
\put(20.5,-4){\makebox[0ex]{\A\p}}
\put(21.3,-3){\makebox[0ex]{\B\p}}
\put(22.7,-3){\makebox[0ex]{\F\c\H}}
\put(23.5,-4){\makebox[0ex]{\G\c\H}}
\put(24.5,-3){\makebox[0ex]{\A\H}}
\put(25.5,-4){\makebox[0ex]{\B\H}}
\put(26.5,-3){\makebox[0ex]{\C\H}}
\put(27.5,-2){\makebox[0ex]{\D\H}}
\put(28.5,-3){\makebox[0ex]{\E\H}}
\put(29.5,-4){\makebox[0ex]{\F\H}}
\put(30.5,-3){\makebox[0ex]{\G\H}}
\put(31.5,-4){\makebox[0ex]{\A\p\H}}
\put(32.5,-3){\makebox[0ex]{\B\p\H}}
\put(1,7){\makebox[0ex]{\F\s\c\V}}
\put(2,6){\makebox[0ex]{\A\f\V}}
\put(3,7){\makebox[0ex]{\B\f\V}}
\put(5,7){\makebox[0ex]{\C\s\V}}
\put(6,6){\makebox[0ex]{\E\f\V}}
\put(8,7){\makebox[0ex]{\F\s\V}}
\put(9,6){\makebox[0ex]{\G\s\V}}
\put(10,7){\makebox[0ex]{\B\f\p\V}}
\put(12,7){\makebox[0ex]{\F\s\c}}
\put(13,6){\makebox[0ex]{\A\f}}
\put(14,7){\makebox[0ex]{\B\f}}
\put(16,7){\makebox[0ex]{\C\s}}
\put(17,6){\makebox[0ex]{\E\f}}
\put(19,7){\makebox[0ex]{\F\s}}
\put(20,6){\makebox[0ex]{\G\s}}
\put(21,7){\makebox[0ex]{\B\f\p}}
\put(23,7){\makebox[0ex]{\F\s\c\H}}
\put(24,6){\makebox[0ex]{\A\f\H}}
\put(25,7){\makebox[0ex]{\B\f\H}}
\put(27,7){\makebox[0ex]{\C\s\H}}
\put(28,6){\makebox[0ex]{\E\f\H}}
\put(30,7){\makebox[0ex]{\F\s\H}}
\put(31,6){\makebox[0ex]{\G\s\H}}
\put(32,7){\makebox[0ex]{\B\f\p\H}}
\end{picture}
\end{center}
  \caption{Ideal Keyboard for Pyth-3 or 19-EDT}
  \label{figure-ideal-keyboard}
\end{figure}

  In our discussion of the equal tempered scale 19-EDT we mentioned already
  that, in principle,
  the scale can be played on the piano with keyboard and tuning
  as usual.
  Yet the keyboard in Figure~\ref{figure-ideal-keyboard}
  better reflects harmonic properties of the scale:
  in particular, the periodicity given by the tritave
  and the assignment of notes with lower harmonic degrees to white keys.  
  Since the harmonic degree $\mu_2$ is invariant under the shift
  by a tritave, the white keys all have notes with harmonic degrees between -5 and 5.
  For each note symbol, the harmonic degree is
  as indicated in Figure~\ref{fig-diatonic-keyboard}.

\FloatBarrier

\subsection{The score of the Ave Maria in dix-neuf par duodecime}
\label{appendix-ave}

For a recording please visit
{\tt\footnotesize https://youtu.be/Bg1n4jM1n5w}.

\begin{center}
  \includegraphics[width=\linewidth]{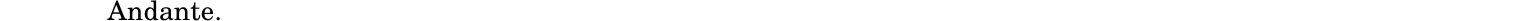}
  
  \includegraphics[width=\linewidth]{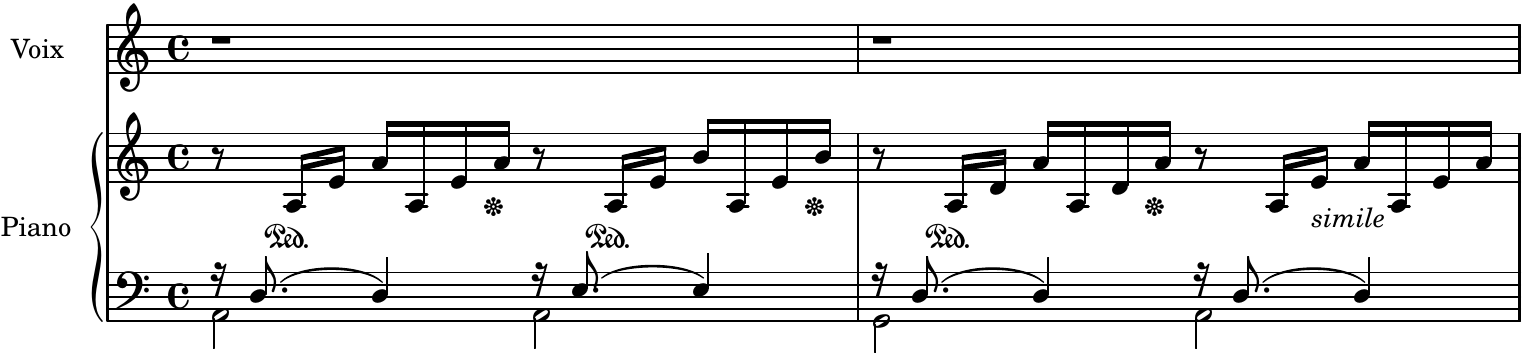}

  \smallskip
  \includegraphics[width=\linewidth]{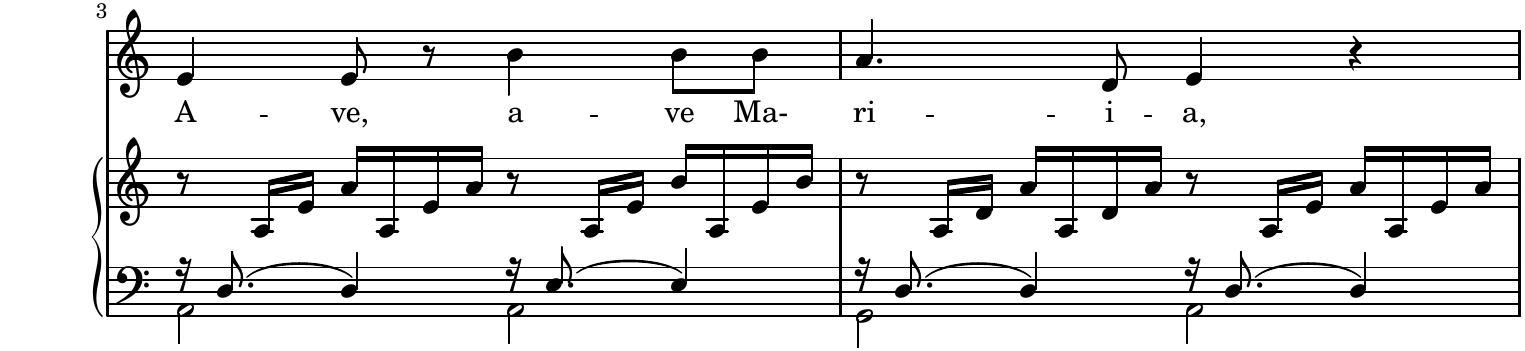}

  \smallskip
  \includegraphics[width=\linewidth]{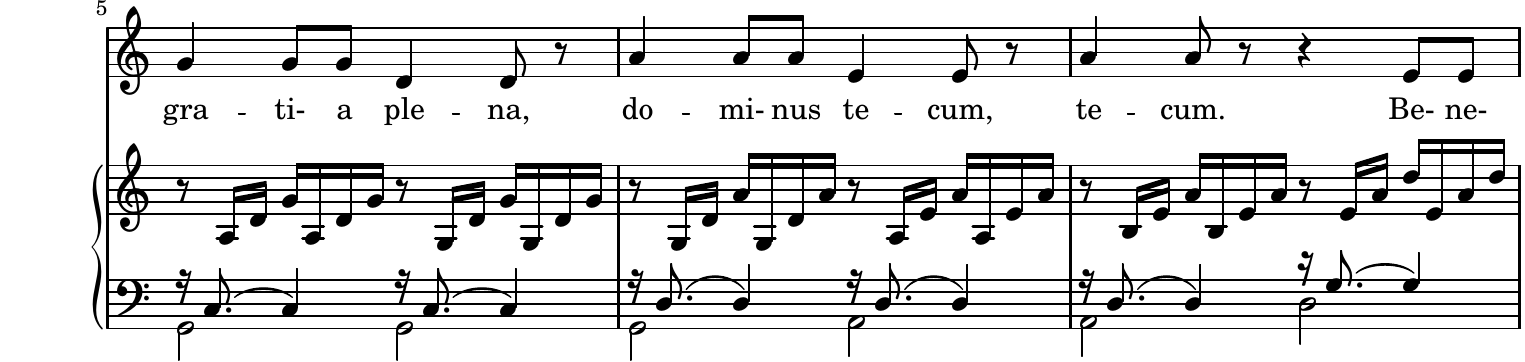}

  \smallskip
  \includegraphics[width=\linewidth]{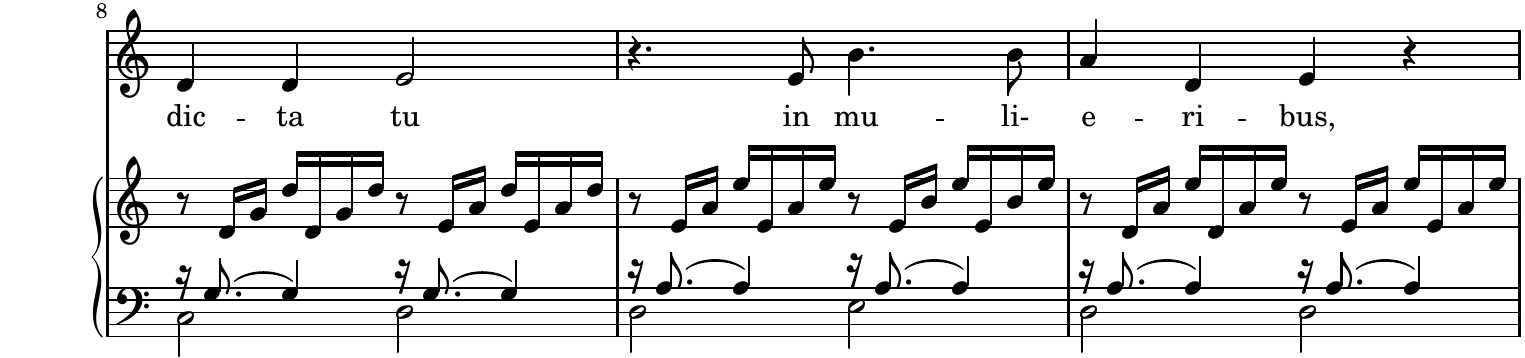}

  \smallskip
  \includegraphics[width=\linewidth]{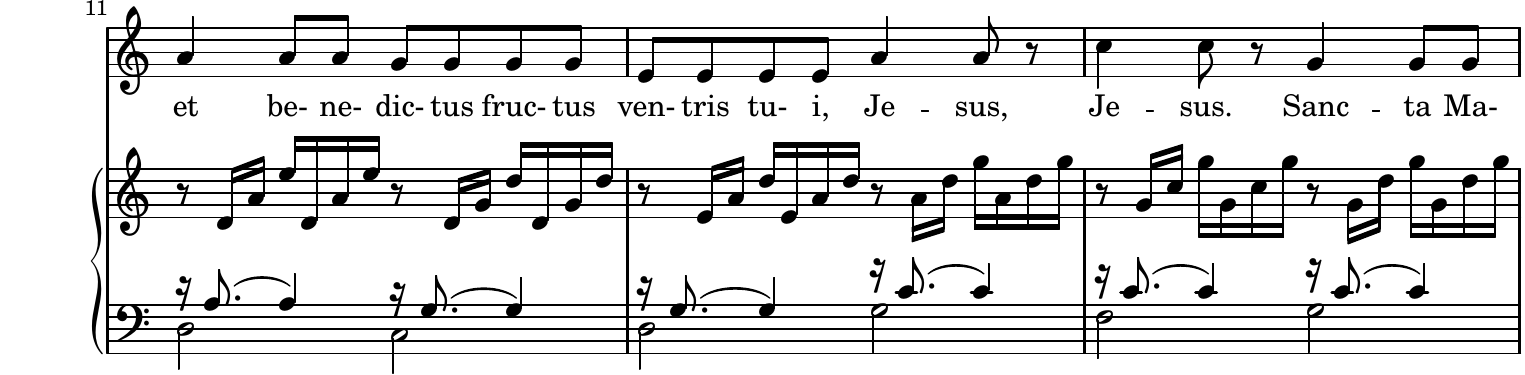}

  \smallskip
  \includegraphics[width=\linewidth]{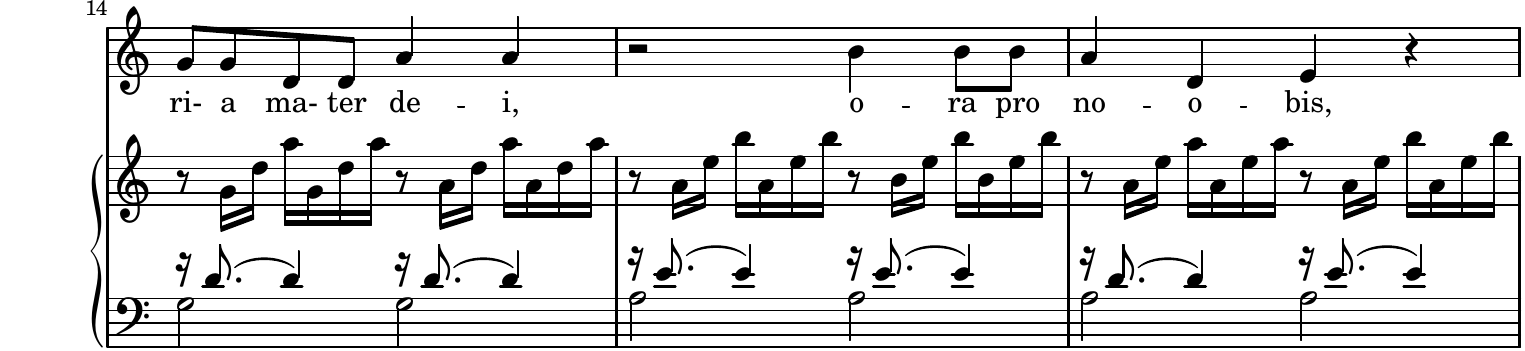}

  \smallskip
  \includegraphics[width=\linewidth]{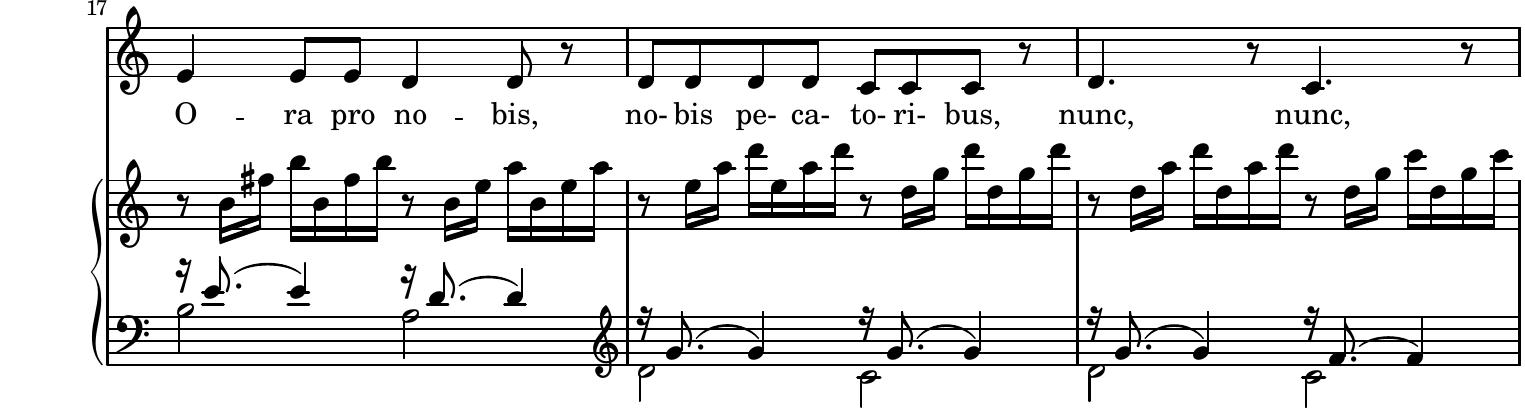}

  \smallskip
  \includegraphics[width=\linewidth]{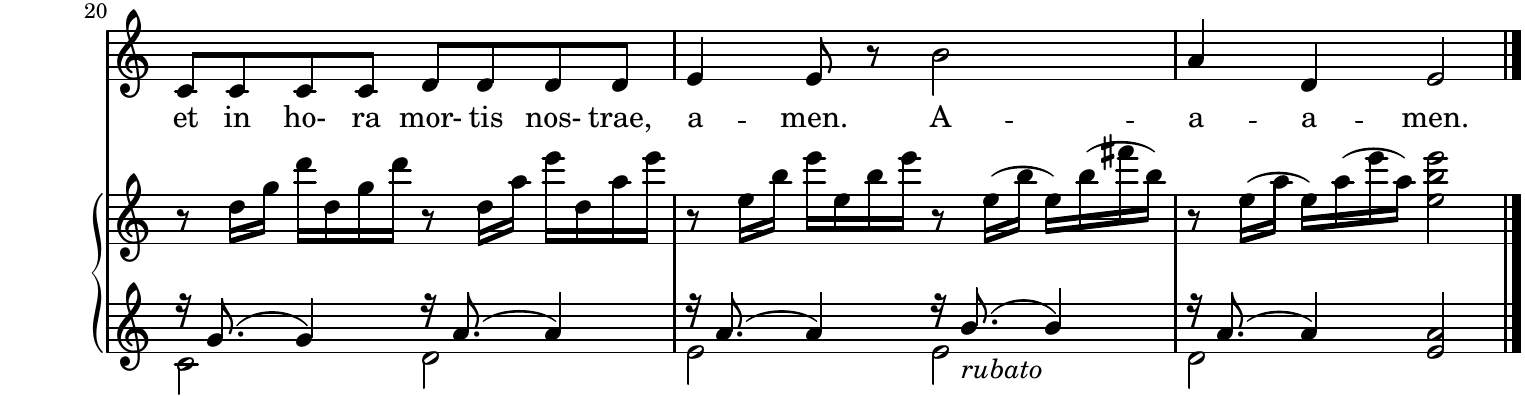}
\end{center}

\FloatBarrier

\subsection{Two tables about chord purity}
\label{subsection-appendix-purity}
We refer to Sections~\ref{section-inversions} and \ref{subsection-purity} for the two measures of
purity exhibited for the chords in the following two tables.

\begin{table}
  \small
\caption{Two measures of purity for 2:3:4 chords}
{\begin{tabular}{|l|c|c|c|c|c|} \hline
\multicolumn6{|c|}{Proximity of common base note and overtone for 2:3:4 chords}
\\ \hline
\multicolumn{1}{|c|}{\it type} & & \multicolumn2{c|}{} & {\it base} & {\it over-} \\ [-1ex]
\multicolumn{1}{|c|}{\it of} & {\it example} &    \multicolumn2{c|}{\it frequency ratios} & {\it note,} & {\it tone,} \\ [-1ex]
\multicolumn{1}{|c|}{\it chord} & & \multicolumn2{c|}{} & {\it distance} $d_B$ & {\it distance} $d_O$ \\ \hline
Major & \A-\E-\A\p\ & $2:3:4$ & $\frac16:\frac14:\frac13$
      & \E\V=\A\c, 2 & \A\p\H=\E\p\p, 3 \\ \hline
Minor & \A-\D-\A\p\ & $3:4:6$ & $\frac14:\frac13:\frac12$
      & \A\V=\D\c\c, 3 & \D\H=\A\p\p, 2 \\ \hline
Augmented & \A-\E-\B\p\ & $4:6:9$ & $\frac19:\frac16:\frac14$
	 & \B\p\V\V=\A\c\c, 4 & \A\H\H=\B\p\p\p, 4 \\ \hline
Diminished & \A-\D-\G\ & $9:12:16$ & $\frac1{16}:\frac1{12}:\frac19$
	   & \A\V\V=\G\c\c\c\c, 9 & \G\H\H=\A\p\p\p\p, 9 \\ \hline
\end{tabular}}
\end{table}

\begin{table}
  \footnotesize
\caption{Two measures of purity for 4:5:6 chords}
{\begin{tabular}{|l|c|c|c|c|c|} \hline
\multicolumn6{|c|}{Proximity of base note and overtone for 4:5:6 chords}
\\ \hline
\multicolumn{1}{|c|}{\it type} & & \multicolumn2{c|}{} & {\it base-} & {\it over-} \\ [-1ex]
\multicolumn{1}{|c|}{\it of} & {\it example} &    \multicolumn2{c|}{\it frequency ratios} & {\it note,} & {\it tone,} \\ [-1ex]
\multicolumn{1}{|c|}{\it chord} & & \multicolumn2{c|}{} & {\it distance} $d_B$ & {\it distance} $d_O$ \\ \hline
Major & \C-\E-\G\ & $4:5:6$ & $\frac1{15}:\frac1{12}:\frac1{10}$
      & \C\c\c, 4 & \B\p\p\p\p, 10 \\ \hline
Major, 1st inv.\ & \E-\G-\C\p\ & $5:6:8$ & $\frac1{24}:\frac1{20}:\frac1{15}$
       & \C\c\c, 5 & \B\p\p\p\p\p, 15 \\ \hline
Major, 2nd inv.\ & \G-\C\p-\E\p\ & $3:4:5$ & $\frac1{20}:\frac1{15}:\frac1{12}$
       & \C\c, 3 & \B\p\p\p\p\p, 12 \\ \hline
Minor & \A-\C-\E\ & $10:12:15$ & $\frac1{6}:\frac1{5}:\frac1{4}$ &
      \F\c\c\c\c, 10 & \E\p\p, 4 \\ \hline
Minor, 1st inv.\ & \C-\E-\A\p\ & $12:15:20$ & $\frac1{5}:\frac1{4}:\frac1{3}$
       & \F\c\c\c\c, 12 & \E\p\p, 3 \\ \hline
Minor, 2nd inv.\  & \E-\A\p-\C\p\ & $15:20:24$ & $\frac1{8}:\frac1{6}:\frac1{5}$
       & \F\c\c\c\c, 15 & \E\p\p\p, 5 \\ \hline
Augmented & \C-\E-\G\s\ & $16:20:25$ & $\frac1{25}:\frac1{20}:\frac1{16}$
	 & \C\c\c\c\c, 16 & \G\s\p\p\p\p, 16 \\ \hline
Diminished & \B-\D-\F\ & $25:30:36$ & $\frac1{36}:\frac1{30}:\frac1{25}$ &
	   \E\f\c\c\c\c\c, 25 & \C\s\p\p\p\p\p, 25 \\ \hline
\end{tabular}}
\end{table}

The {\it common base note} of a chord is the highest note such that
each note in the chord is an overtone;  its {\it distance} $d_B$ from the
lowest note in the chord is given by the first entry in the integer
frequency ratios.
The {\it common over tone} of a chord is the lowest note which occurs
as an overtone for each note in the chord.  Its {\it distance} $d_O$ from the
highest note in the chord can be read off from the last entry in the
reciprocal frequency ratios.

\bigskip
 Address of the author:
 \nopagebreak
 \parbox[t]{5.5cm}{\footnotesize
              Department of\\
              Mathematical Sciences\\ 
              Florida Atlantic University\\
              777 Glades Road\\
              Boca Raton, Florida 33431\\[1ex]
	      {\tt markusschmidmeier@gmail.com}
	      }     

\end{document}